\documentclass{aa}  

\usepackage{graphicx}
\usepackage{txfonts,textcomp}
\usepackage{gensymb} 
\usepackage{xspace}
\usepackage{tabularx}
\usepackage{xcolor}
\usepackage{placeins}
\usepackage{threeparttable}
\usepackage{ulem}
\usepackage{natbib,twoopt}
\usepackage[breaklinks=true]{hyperref} 
\bibpunct{(}{)}{;}{a}{}{,} 
\makeatletter
\newcommandtwoopt{\citeads}[3][][]{\href{http://adsabs.harvard.edu/abs/#3}%
{\def\hyper@linkstart##1##2{}%
\let\hyper@linkend\@empty\citealp[#1][#2]{#3}}}
\newcommandtwoopt{\citepads}[3][][]{\href{http://adsabs.harvard.edu/abs/#3}%
{\def\hyper@linkstart##1##2{}
\let\hyper@linkend\@empty\citep[#1][#2]{#3}}}
\newcommandtwoopt{\citetads}[3][][]{\href{http://adsabs.harvard.edu/abs/#3}%
{\def\hyper@linkstart##1##2{}
\let\hyper@linkend\@empty\citet[#1][#2]{#3}}}
\newcommandtwoopt{\citeyearads}[3][][]%
{\href{http://adsabs.harvard.edu/abs/#3}
{\def\hyper@linkstart##1##2{}%
\let\hyper@linkend\@empty\citeyear[#1][#2]{#3}}}
\makeatother
\def\ms{\hbox{m\,s$^{-1}$}\xspace}         
\def\m2s2{\hbox{\,m$^{2}$\,s$^{-2}$}\xspace} 
\def\kms{\hbox{\,km\,s$^{-1}$}\xspace}       
\def\vsini{\hbox{$v$\,sin\,$i_{\star}$}\xspace}      
\def\Msun{$M_{\odot}$\xspace}             
\def\Rsun{$R_{\odot}$\xspace}
\def\Mjup{\hbox{$\mathrm{M}_{\rm J}$}\xspace}

\def\ten[#1]{$\;\times 10^{#1}$}

\def\logg{$\log g$}

\newcommand{\e}[1]{{\times10^{#1}}}

\newcommand{\Rnom}{\hbox{$\mathcal{R}^{\rm N}_{\odot}$}} 

\newcommand{\GMnom}{\hbox{$\mathcal{(GM)}^{\rm N}_{\odot}$}}

\newcommand{\Renom}{\hbox{$\mathcal{R}^{\rm N}_{e \rm E}$}}

\newcommand{\GMenom}{\hbox{$\mathcal{(GM)}^{\rm N}_{\rm E}$}}

\newcommand{\RJnom}{\hbox{$\mathcal{R}^{\rm N}_{e \rm J}$}}

\newcommand{\GMJnom}{\hbox{$\mathcal{(GM)}^{\rm N}_{\rm J}$}}

\newcommand{\rebound}{{\sc \tt REBOUND}\xspace}
\newcommand{\whf}{{\sc \tt WHFast}\xspace}
\newcommand{\emcee}{{\sc \tt emcee}\xspace}
\newcommand{\juliet}{{\sc \tt juliet}\xspace}
\newcommand{\batman}{{\sc \tt batman}\xspace}

\newcommand{\celerite}{{\sc \tt celerite}\xspace}

\newcommand{\PARSEC}{{\sc \tt PARSEC}\xspace}

\newcommand{\gcm}{$\mathrm{g\;cm^{-3}}$}

\newcommand{\pipe}{{\sc \tt PIPE}\xspace}
\newcommand{\astropy}{{\sc \tt astropy}\xspace}
\newcommand{\prose}{{\sc \tt prose}\xspace}
\newcommand{\photutils}{{\sc \tt photutils}\xspace}

\newcommand{\nuance}{{\sc \tt nuance}\xspace}

\newcommand{\gastli}{{\sc \tt GASTLI}\xspace}

\newcommand{\tesscont}{{\sc \tt TESS-cont}\xspace}

\newcommand{\REarth}{$\mathrm{R_E}$\xspace}
\newcommand{\MEarth}{$\mathrm{M_E}$\xspace}

\newcommand{\be}{\begin{equation}}
\newcommand{\ee}{\end{equation}}

\newcommand{\rn}[1]{(\ref{#1})}

\newcommand{\bea}{\begin{eqnarray}}
\newcommand{\eea}{\end{eqnarray}}
\newcommand{\ff}[2]{{\textstyle \frac{#1}{#2}}}

\newcommand{\ben}{\begin{enumerate}}
\newcommand{\een}{\end{enumerate}}

\def\logg{$\log g$}
\def\Msun{$M_{\odot}$\xspace}            
\def\Rsun{$R_{\odot}$\xspace}
\newcommand{\orcid}[1]{\protect\href{https://orcid.org/#1}{\protect\includegraphics[width=8pt]{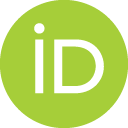}}}
\hypersetup{colorlinks=true, citecolor=blue, linkcolor=blue}

\begin{document}


    \title{HD\,148797: A bright F-type star with two moderate-period low-density sub-Jovian planets\thanks{This study uses CHEOPS data observed as part of the Discretionary Programme (DP) PR460045 (PI Almenara).}}
    
    \subtitle{Compact multi-planet architectures are common in the Neptunian savanna}

    \titlerunning{HD\,148797}
    
%
%
%

    \author{J.M.~Almenara\orcid{0000-0003-3208-9815}\inst{\ref{Geneva}}
        \and A.~Castro-Gonz\'alez\orcid{0000-0001-7439-3618}\inst{\ref{Geneva}}
        \and R.~Mardling\inst{\ref{Monash},\ref{Geneva}} 
        \and T.~Guillot\orcid{0000-0002-7188-8428}\inst{\ref{Nice}}
        \and X.~Bonfils\orcid{0000-0001-9003-8894}\inst{\ref{Grenoble}}
        \and L.~Thomas\orcid{0009-0006-1571-0306}\inst{\ref{Munich},\ref{Garching}}
        \and L.~Gauvrit\inst{\ref{Nice}}
        \and L.~Abe\orcid{0000-0002-0856-4527}\inst{\ref{Nice}}
        \and A.~Agabi\inst{\ref{Nice}}
        \and J.~Aubert\orcid{0009-0003-4077-473X}\inst{\ref{Geneva}} 
        \and F.~Bouchy\orcid{0000-0002-7613-393X}\inst{\ref{Geneva}}
        \and C.~Cadieux\inst{\ref{Geneva}}
        \and A.~Deline\inst{\ref{Geneva}} 
        \and R.F.~D\'{i}az\inst{\ref{ITBA},\ref{BA}}
        \and D.~Ehrenreich\orcid{0000-0001-9704-5405}\inst{\ref{Geneva}} 
        \and T.~Forveille\inst{\ref{Grenoble}}
        \and A.~Heitzmann\inst{\ref{Geneva}} 
        \and P.~Hirling\inst{\ref{Geneva}} 
        \and R.M.~Hoogenboom\orcid{0009-0004-9519-9143}\inst{\ref{Geneva}} 
        \and A.~Leleu\inst{\ref{Geneva}} 
        \and M.~Lendl\orcid{0000-0001-9699-1459}\inst{\ref{Geneva}} 
        \and D.~M\'{e}karnia\orcid{0000-0001-5000-7292}\inst{\ref{Nice}}
        \and S.~Pelletier\inst{\ref{Geneva}}
        \and A.~Psaridi\orcid{0000-0002-4797-2419}\inst{\ref{Barcelona1},\ref{Barcelona2}}
        \and E.~Rea\orcid{0009-0009-4849-9764}\inst{\ref{ISP},\ref{PNRA}}
        \and X.~Song\inst{\ref{Geneva},\ref{Beijing}}
        \and O.~Su\'{a}rez\orcid{0000-0002-3503-3617}\inst{\ref{Nice}}
        \and A.H.M.J.~Triaud\orcid{0000-0002-5510-8751}\inst{\ref{Birmingham}}
        \and S.~Udry\inst{\ref{Geneva}}          
        }

   \institute{Observatoire de Gen\`eve, Département d’Astronomie, Universit\'e de Gen\`eve, Chemin Pegasi 51b, 1290 Versoix, Switzerland\label{Geneva}
        \and School of Physics and Astronomy, Monash University, Victoria, 3800, Australia\label{Monash}
        \and Universit\'e C\^ote d'Azur, Laboratoire Lagrange, OCA, CNRS UMR 7293, Nice, France\label{Nice}
        \and Univ. Grenoble Alpes, CNRS, IPAG, F-38000 Grenoble, France\label{Grenoble}
        \and University Observatory Munich, Faculty of Physics, Ludwig-Maximilians-Universit\"at München, Scheinerstr. 1, 81679 Munich, Germany\label{Munich}
        \and Max-Planck Institute for Extraterrestrial Physics, Giessenbachstrasse 1, D-85748 Garching, Germany\label{Garching}
        \and Instituto Tecnol\'ogico de Buenos Aires (ITBA), Iguaz\'u 341, Buenos Aires, CABA C1437, Argentina\label{ITBA}
        \and Instituto de Ciencias F\'isicas (ICIFI; CONICET), ECyT-UNSAM, Campus Miguelete, 25 de Mayo y Francia, (1650) Buenos Aires, Argentina\label{BA}
        \and Institute of Space Sciences (ICE, CSIC), Carrer de Can Magrans S/N, Campus UAB, Cerdanyola del Valles, E-08193, Spain\label{Barcelona1}
        \and Institut d’Estudis Espacials de Catalunya (IEEC), 08860 Castelldefels (Barcelona), Spain\label{Barcelona2}
        \and Istituto di Scienze Polari del CNR (ISP-CNR), Università Ca’ Foscari, Via Torino n. 155, 30172 Venezia Mestre (VE), Italy\label{ISP}
        \and Programma Nazionale di Ricerche in Antartide (PNRA), Institut polaire français Paul-\'Emile Victor (IPEV)\label{PNRA}        
        \and Laboratory for Climate and Ocean-Atmosphere Studies, Department of Atmospheric and Oceanic Sciences, School of Physics, Peking University, Beijing 100871, China\label{Beijing}
        \and School of Physics \& Astronomy, University of Birmingham, Edgbaston, Birmingham B15 2TT, UK\label{Birmingham}
    }
   
   \date{}

 
  \abstract
{
    We report the confirmation and characterisation of two moderate-period sub-Jovian planets transiting the bright F-type star HD\,148797 ($G=9.4$~mag, $T_{\rm eff}=6441\pm51$~K). Using photometric time series from TESS and CHEOPS, we determine orbital periods of 42.1~d for HD\,148797~b and 68.2~d for HD\,148797~c, putting the period ratio very close to the golden mean at 1.619 and therefore near several strong harmonics, and measure planetary radii of $8.25 \pm 0.37$~\REarth and $8.37 \pm 0.38$~\REarth, respectively. We detect significant anti-correlated transit-timing variations for both planets, which contain enough harmonic information to yield photodynamical masses of $39.3^{+13}_{-8.5}$~\MEarth for HD\,148797~b and $39.6 \pm 9.3$~\MEarth for HD\,148797~c. The corresponding bulk densities, $0.39 \pm 0.12$ and $0.377 \pm 0.084$~\gcm, place both planets among the low-density sub-Jovians of the Neptunian savanna. The architecture of HD\,148797 is not unusual within this regime: we find that detected multi-system fractions in the savanna remain at $\sim70$--$90\%$, and that most savanna multi-planet systems contain at least one adjacent planet pair with $P_{\rm out}/P_{\rm in}<3$. This pattern suggests that savanna sub-Jovians are commonly found in dynamically cold systems, consistent with smoother migration pathways such as disk-driven migration rather than disruptive high-eccentricity tidal migration. As a bright, co-evolved system hosting two warm savanna sub-Jovians with similar radii and masses, HD\,148797 is also a promising target for comparative atmospheric characterisation.
}
   \keywords{stars: individual: \object{HD\,148797} --
        stars: planetary systems --
        techniques: photometric
               }

   \maketitle

\section{Introduction}\label{section:introduction}

The orbital distribution of giant planets (i.e. $R_{\rm p} \gtrsim 4$~\REarth) carries important information about their formation and migration histories. Giant planets are thought to form most efficiently at large orbital distances, beyond the ice line, where the solid material in the protoplanetary disk can build planetary cores massive enough to trigger runaway gas accretion ($\sim$10~\MEarth; \citealt[][]{Rafikov2006,Lee2015,Lee2019}). Their presence at shorter orbital periods therefore generally implies substantial inward migration. A well-known example is the hot-Jupiter pile-up at periods of a few days \citep[e.g.][]{Udry2003,Udry2007}, commonly interpreted as the outcome of high-eccentricity tidal migration \citep[HEM; e.g.][]{Wu2003,Correia2011,2026arXiv260620789Z}. Recently, an analogous overdensity has been identified in the sub-Jovian population at $P_{\rm orb}\simeq3$--$6$~d, the so-called Neptunian ridge \citep{CastroGonzalez2024a}. This similarity, together with recent theoretical and observational results, suggests that the ridge may also be largely populated by planets that underwent HEM \citep[e.g.][]{Bourrier2018,Bourrier2023,Correia2020,2025AJ....169..117V,2025AJ....170..275Y,CastroGonzalez2026}. In contrast, the Neptunian savanna \citep{Bourrier2023}, where sub-Jovians are more sparsely distributed ($P_{\rm orb} \gtrsim 6$~d; \citealt{CastroGonzalez2024a}) and tidal dissipation is less efficient \citep[e.g.][]{Hut1980}, may trace smoother migration pathways, such as disk-driven migration \citep[DDM; e.g.][]{Goldreich1979,Lin1996}.

The present-day orbital architectures of close-in giant planets provide an additional diagnostic of their migration histories. In standard HEM scenarios, a planet is driven onto a highly eccentric orbit by a distant massive planetary or stellar companion and subsequently circularized by tides at short orbital periods \citep[e.g.][]{Wu2003,Fabrycky2007}. This process is therefore expected to leave an inner short-period giant planet and a more distant perturber, with few, if any, additional planets at intermediate orbital periods. This expectation is supported by numerical simulations showing that HEM typically destabilizes and removes lower-mass inner planets during the tidal circularisation phase \citep[e.g.][]{Mustill2015}. In contrast, DDM does not require a distant eccentricity-exciting companion and proceeds through a smoother interaction with the protoplanetary disk, making it more compatible with the survival of compact multi-planet systems. This picture is consistent with the observed tendency for hot Jupiters in the pile-up to be more isolated than warm Jupiters \citep[e.g.][]{Huang2016}. However, whether a similar architectural dichotomy exists between sub-Jovians in the ridge and savanna has remained unexplored.

Planetary bulk densities add an important physical dimension to this picture. In the hot-Jupiter pile-up, the interpretation of bulk densities is complicated by radius inflation, which affects strongly irradiated giant planets and obscures the connection between their present-day densities and their primordial internal structures \citep[e.g.][]{Fortney2010,Demory2011}. However, such highly inflated planets are not observed in the ridge. \citet{CastroGonzalez2024b} found that planets in the savanna typically have low densities ($\rho_{\rm p}\lesssim1~\mathrm{g}~\mathrm{cm}^{-3}$), whereas planets in the ridge can reach substantially higher densities, up to $\sim2.5~\mathrm{g}~\mathrm{cm}^{-3}$. Although ridge planets are strongly irradiated, objects in this sub-Jovian mass and radius regime are expected to be largely stable against photoevaporative mass loss \citep[e.g.][]{Ionov2018,Vissapragada2022}. The enhanced densities observed in the ridge are therefore unlikely to be produced by atmospheric escape. Instead, they may reflect intrinsically denser internal structures among planets that migrated through HEM \citep{CastroGonzalez2024b}, or the partial tidal stripping of initially low-density planets during HEM \citep{CastroGonzalez2026,2026ApJ...997..138H}. Nevertheless, this density trend is well established only up to $P_{\rm orb}\lesssim30$~d. Extending density measurements to wider orbits is therefore essential: the discovery of dense sub-Jovians at larger separations, closer to the regions where such planets formed, would support the idea that some ridge planets were already dense before migrating inward, whereas a persistently low-density savanna would favour post-formation evolutionary processes, such as tidal stripping, as the origin of the enhanced densities observed in the ridge.

\citet{Salinas2025} flagged HD\,148797 (TIC~221567884) as a candidate multi-planet system in a machine-learning search of Transiting Exoplanet Survey Satellite \citep[TESS;][]{Ricker2015} light curves, based on multiple transit-like events observed across several TESS sectors. Here, we confirm two transiting sub-Jovian planets orbiting HD\,148797 with periods of 42.1~d and 68.2~d, placing both planets in the warm, moderate-period Neptunian savanna. The host is a bright ($G=9.4$~mag), moderately rotating F-type star with $T_{\rm eff}=6441\pm51$~K and \vsini$\sim$20~\kms, close to the Kraft break \citep{Kraft1967,Beyer2024}. Its broad stellar spectral lines make precise radial velocity (RV) mass measurements challenging. However, the planets exhibit anticorrelated transit-timing variations \citep[TTVs;][]{Agol2005,Holman2005}, which dynamically link the two signals to the same system and allow us to measure their masses and bulk densities. HD\,148797 therefore provides a valuable opportunity to characterize two moderate-period savanna sub-Jovians in the same system, extending density and architecture measurements into a regime of the savanna that has been comparatively less explored.

Section~\ref{section:observations} presents the TESS, ground-based, and CHEOPS photometric observations. Section~\ref{section:stellar_parameters} describes the characterisation of the host star. Section~\ref{section:analysis_results} presents the photodynamical analysis and the resulting system parameters. Section~\ref{section:discussion} places HD\,148797 in the context of similar Neptunian savanna systems and discusses its dynamics, internal structure, and prospects for atmospheric characterisation. Finally, Sect.~\ref{section:conclusions} summarizes our main conclusions.

\section{Observations}\label{section:observations}

\begin{table*}[!htbp]
\renewcommand{\arraystretch}{1.1}
\small
\caption{Summary of the CHEOPS observations of HD\,148797.}          
\label{table:cheops}      
\centering                          
\begin{tabular}{l c c c c c}        
\hline\hline                 
File key & OBS ID & UTC start & Visit duration & Exposure time & Efficiency \\    
\hline                        
\verb|CH_PR460045_TG000201_V0300| & 3130520 & 2026-06-21T03:41:19 & 44841 s & 1 x 60.0 s & 65.7\% \\
\verb|CH_PR460045_TG000101_V0300| & 3131328 & 2026-06-25T01:14:48 & 38778 s & 1 x 60.0 s & 65.8\% \\
\hline                                   
\end{tabular}
\end{table*}

\subsection{TESS}

TESS \citep{Ricker2015} observed HD\,148797 in full-frame images (FFIs) during Sectors~12, 39, 66, and~93, with cadences of 1800, 600, 200, and 200~s, respectively. The target was later observed in target-pixel (TP) mode in Sectors~102 and~103 with a cadence of 120~s. We built a homogeneous photometric dataset following the Simple Aperture Photometry (SAP) approach used by the TESS Science Processing Operations Center \citep[SPOC;][]{Jenkins2016,Caldwell2020}. For each sector, we used \tesscont\footnote{\url{https://github.com/castro-gzlz/TESS-cont}} \citep{CastroGonzalez2024b} to define custom apertures optimized to minimize flux contamination. The remaining dilution was then corrected using the FLFRCSAP and CROWDSAP metrics recomputed for each custom aperture, following \citet{CastroGonzalez2020,CastroGonzalez2022}. Figure~\ref{fig:TESS-cont} shows a representative target-pixel-file (TPF)-shaped heatmap of the target-flux fraction in each pixel, together with the selected aperture and the main contaminating sources. For all sectors, we find contaminant flux fractions of $\lesssim 0.5\%$. For the photodynamical analysis of Sect.~\ref{section:analysis_results}, we construct a dataset consisting of 24-hour segments centred on the transits of planets~b and c.

The period of planet~c was unambiguously determined only after the TESS Sector~102 observations. Before that, 17 period aliases remained possible, and we used EulerCam \citep[ECAM;][]{Lendl2012} and the Antarctica Search for Transiting ExoPlanets \citep[ASTEP;][]{Guillot2015,Mekarnia2016} to test and rule out some of them. Once the period was determined by TESS, targeted follow-up transits were obtained with the CHaracterising ExOPlanet Satellite \citep[CHEOPS;][]{Benz2021} and ASTEP to add new transits to the TTV analysis.

\subsection{EulerCam}

We used EulerCam \citep[ECAM;][]{Lendl2012} on the Swiss 1.2-m Euler telescope at La Silla Observatory to observe two possible transit windows of planet~c on the nights of 21~March and 8~April 2026. These windows corresponded to aliases~3, 6, 9, 12, and~15 on 21~March, and alias~14 on 8~April.\footnote{Alias number~1 corresponds to the maximum allowed period ($\sim$750~d), and subsequent alias numbers correspond to integer subdivisions of that period.} No transit-like signal was detected in these windows, allowing us to rule out the corresponding period aliases. The observations were conducted with the $r'$-Gunn filter and an exposure time of 30~s. The image reduction was carried out following \citet{Lendl2012} and \citet{Lendl2014}. Aperture photometry was performed with \prose \citep{Garcia2022}, which relies on \astropy \citep{astropy2022} and \photutils \citep{Bradley2023}, and differential photometry was performed following the approach of \citet{Broeg2005}.

\subsection{ASTEP}

We used ASTEP \citep{Guillot2015,Mekarnia2016}, a 0.4-m telescope located at Dome~C on the east Antarctic plateau. ASTEP is equipped with a Wynne-Newtonian coma corrector and a focal box hosting two high-sensitivity cameras operating at red and blue wavelengths \citep{Schmider2022,Dransfield2022}. We obtained additional coverage of the transit window of planet~c on the night of 22~March 2026, corresponding to aliases~3, 6, 9, 12, and~15. These data provided an independent non-detection of the same transit window. In addition, ASTEP observed the egress of planet c on 21~June 2026 and that of planet~b on 25~June 2026, confirming that both transits occur on target. The observation on 21~June~2026 was obtained through thick clouds, and the light curve from the blue camera shows large systematics. Therefore, only the red‑camera light curve was used. On both nights, exposure times of 60~s (blue camera) and 3~s (red camera) were used. To reduce the number of data points, the red‑camera light curve was binned to 60~s.

\subsection{CHEOPS}

We observed HD\,148797 with CHEOPS \citep{Benz2021} as part of the Programme PR460045 (PI Almenara). The observations targeted the predicted transit windows of planet~c on 21~June~2026 and planet~b on 25~June~2026 (Table~\ref{table:cheops}). As expected for a low-Earth-orbit mission, the light curves contain regular gaps due to Earth occultations, passages through or near the South Atlantic Anomaly, and intervals of enhanced stray light. These interruptions did not compromise the out-of-transit baseline around either event, and the two visits provided late-epoch timing measurements for the TTV analysis. The raw data were automatically processed with the CHEOPS Data Reduction Pipeline \citep[DRP; version 15.1.1;][]{Hoyer2020}. We extracted the final light curves using the \pipe package\footnote{\url{https://github.com/alphapsa/PIPE}} \citep{Brandeker2024}, which performs point-spread-function photometry. The resulting light curves were then included in the joint photodynamical analysis described in Sect.~\ref{section:analysis_results}, where the transit and instrumental models were fitted simultaneously.

\section{Stellar parameters}\label{section:stellar_parameters}

\begin{table}[htb!]
\renewcommand{\arraystretch}{1.1}
    \tiny
      \caption{Stellar properties of HD\,148797.}\label{table:stellar_parameters}
      \centering
        \setlength{\tabcolsep}{4pt}
    \begin{tabular}{lcl}
    \hline
    \hline
    Parameter & Value & Ref. \\
    \hline
    \textit{Designations} & TYC\,8720-1711-1               & 1 \\
                          & 2MASS\,J16333911-5641563      & 2 \\
                          & TIC\,221567884                & 3 \\
                          & Gaia\,DR3\,5928516226857454464 & 4 \smallskip\\

    \textit{Astrometry} \\
    Right ascension (ICRS, J2016), $\alpha$ & 16$^{\rm h}$33$^{\rm m}$39.06$^{\rm s}$ & 4 \\
    Declination (ICRS, J2016), $\delta$ & $-56^{\rm o}$41'57.43'' & 4 \\
    Proper motion $\alpha$ (mas/year) & $-32.599 \pm 0.016$ & 4 \\
    Proper motion $\delta$ (mas/year) & $-60.670 \pm 0.016$ & 4 \\
    Parallax, $\pi$ (mas) & $5.822 \pm 0.016$ & 4, 5 \\
    Distance, d (pc) & $171.77 \pm 0.46$ & $\pi$ \smallskip\\
    
    \textit{Photometry} \\
    \textit{\textit{Gaia}}-BP (mag)   & 9.6829 $\pm$ 0.0029  & 4 \\
    \textit{Gaia}-G (mag)    & 9.4400 $\pm$ 0.0028  & 4 \\
    \textit{Gaia}-RP (mag)   & 9.0373 $\pm$ 0.0028  & 4 \\
    2MASS-J (mag)   &  8.614 $\pm$ 0.024  & 2 \\
    2MASS-H (mag)   &  8.427 $\pm$ 0.051 & 2 \\ 
    2MASS-Ks (mag)  &  8.355 $\pm$ 0.029 & 2 \\
    WISE-W1 (mag)   &  8.312 $\pm$ 0.022 & 6 \\
    WISE-W2 (mag)   &  8.331 $\pm$ 0.019 & 6 \\
    WISE-W3 (mag)   &  8.397 $\pm$ 0.024 & 6 \smallskip\\
    
    \textit{Stellar parameters} \\
    Spectral type & F6 & 7 \\
    Effective temperature, $T_{\rm eff}$ (K) & $6441 \pm 51$   & 8 \\
    Surface gravity, log g (cgs)             & $4.050 \pm 0.077$ & 8 \\
    Metallicity, [M/H] (dex)                 & $-0.071 \pm 0.079$ & 8 \\ 
    Prior stellar mass, $M_\star$ (\Msun)    & $1.268 \pm 0.080$ & 9 \\
    Prior stellar radius, $R_\star$ (\Rsun)  & $1.569 \pm 0.069$ & 9 \\
    \hline
    \end{tabular}
    \begin{tablenotes}
        \tiny
        \item Refs: 1)~\citet{Hog2000}, 2)~\cite{Cutri2003}, 3)~\citet{Stassun2019}, 4)~\cite{GaiaDR3}, 5)~\cite{Lindegren2021}, 6)~\cite{Cutri2013}, 7)~\cite{Pecaut2013}, 8)~\cite{Turchi2025}, 9)~This work.
    \end{tablenotes}
\end{table}

\begin{figure*}[!ht]
  \centering
  \includegraphics[width=0.98\textwidth]{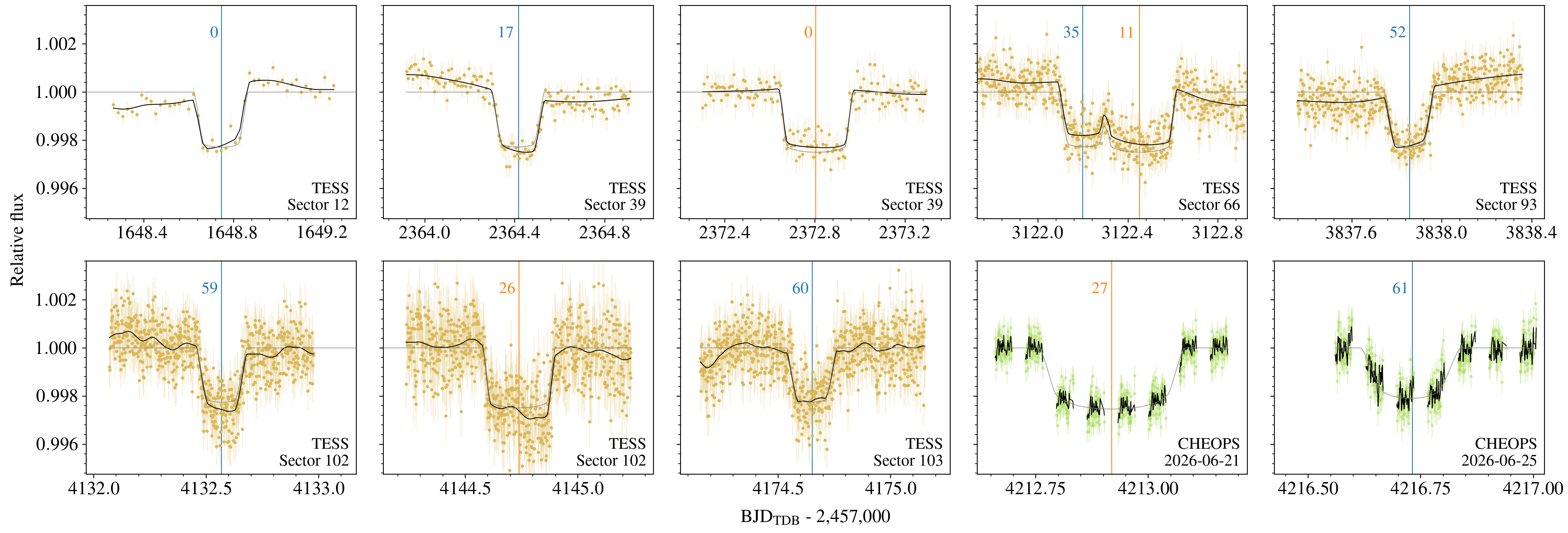}
  \caption{Photodynamical modelling of the space-based transit photometry. The black line is the MAP model that combines both transits and noise, while the gray line shows the pure transit model. Vertical lines mark the midtransit time of planets~b (blue) and c (orange), and are labelled by the number of orbital periods since the first observed transit.} \label{figure:phot}
\end{figure*}

\begin{figure}[!ht]
  \centering
  \includegraphics[width=0.49\textwidth]{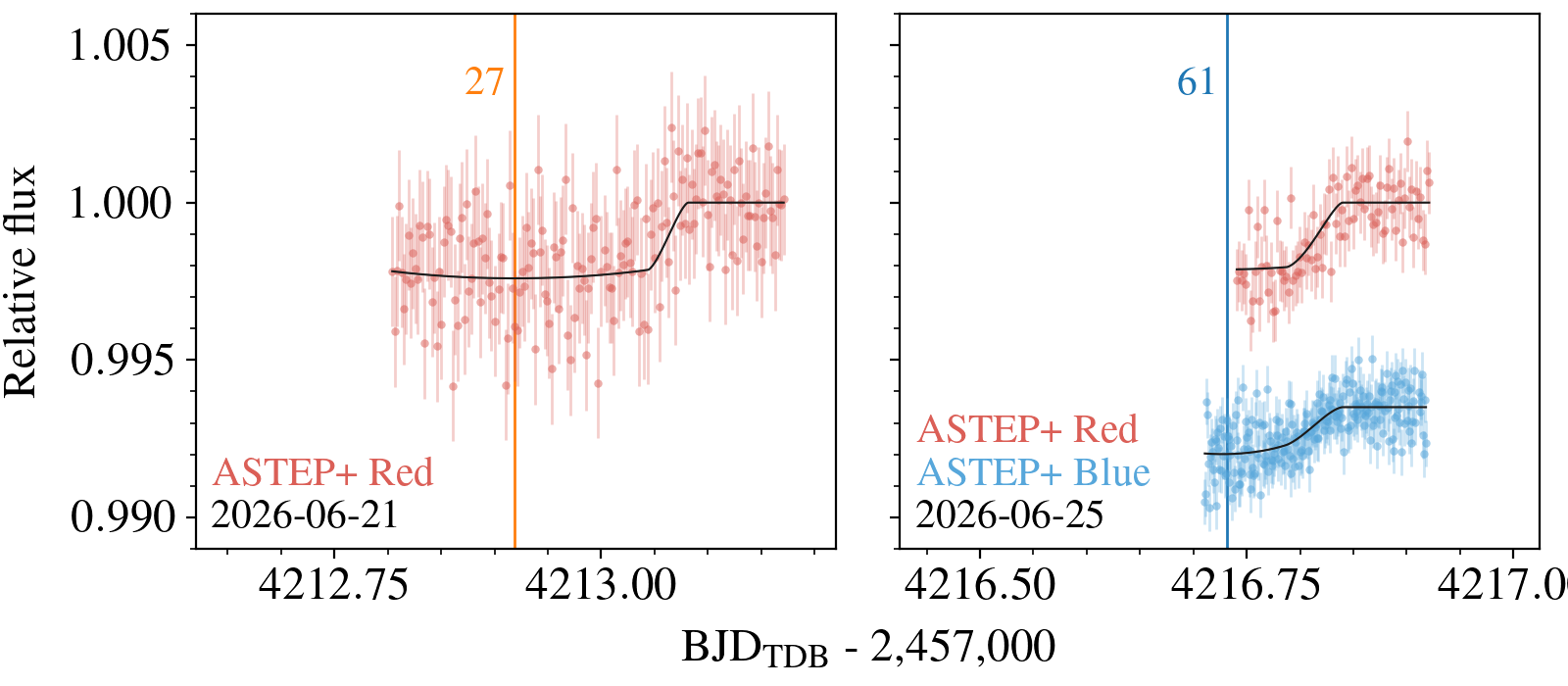}
  \caption{Same as Fig.~\ref{figure:phot}, but for the noise-model-corrected ASTEP observations.} \label{figure:ASTEP}
\end{figure}

To determine the stellar parameters of HD\,148797, we fit its spectral energy distribution (SED) using stellar atmosphere and evolution models, adopting priors on the distance from the \textit{Gaia} parallax \citep{Gaia, GaiaDR3} and on atmospheric parameters from \textit{Gaia} XP spectra \citep{Andrae2023a}. The SED was constructed using photometric data from \textit{Gaia} Data Release 3 \citep[DR3;][]{Riello2021}, the 2-Micron All-Sky Survey \citep[2MASS;][]{2mass,Cutri2003}, and the Wide-field Infrared Survey Explorer \citep[WISE;][]{wise,Cutri2013}. The corresponding measurements are listed in Table~\ref{table:stellar_parameters}. 
We adopted the PHOENIX/BT-Settl atmosphere model \citep{Allard2012}, along with two sets of stellar evolution models: Dartmouth \citep{Dotter2008} and \PARSEC \citep{Chen2014}. The SED fitting followed the procedure described by \citet{Diaz2014}, using informative priors for the effective temperature, surface gravity, and metallicity, from \citet{Turchi2025}. The distance prior was based on the \textit{Gaia} zeropoint-corrected parallax \citep{Lindegren2021}. Uniform priors were adopted for the remaining parameters. We included a jitter term \citep{Gregory2005} for each photometric band set (\textit{Gaia}, 2MASS, and WISE). 
We merged the results from the two models assuming equal probability for each. The priors and posteriors for all parameters are listed in Table~\ref{table:SED}, and the maximum a posteriori (MAP) stellar atmosphere model is shown in Fig.~\ref{figure:SED}. Following \citet{Tayar2022}, we added a systematic uncertainty floor of 4.2\% for the radius and 5\% for the mass, combined in quadrature with the statistical uncertainties (Table~\ref{table:stellar_parameters}). The derived stellar mass and radius ($1.268 \pm 0.080$~\Msun, $1.569 \pm 0.069$~\Rsun) were used as priors in the photodynamical modelling (Sect.~\ref{section:analysis_results}). HD\,148797 has a Gaia DR3 renormalised unit weight error (RUWE) value of 0.876, consistent with a single-star astrometric solution \citep{Castro-Ginard2024}. 

From the Gaia DR3 spectral line broadening ($20.5 \pm 1.7$~\kms) and a macroturbulence estimate ($6.42 \pm 0.30$~\kms) from \citet{Doyle2014}, we infer a $\vsini\,$$\simeq$$\,19.5 \pm 1.8$~\kms, implying the star is a moderate rotator. To estimate the stellar rotational period, we analyse the presearch data-conditioning simple aperture photometry (PDCSAP; \citealt{Smith2012}, \citealt{Stumpe2012,Stumpe2014}) light curves of TESS Sectors 12, 39, 66, 102 and 103 with a rotation kernel \citep{Foreman-Mackey2017} Gaussian process (GP) using \juliet \citep{Espinoza2019,Kreidberg2015,Speagle2020}. Sector 39 shows the clearest hint of a periodic modulation (Fig.~\ref{figure:Prot}), from which we tentatively infer a rotational period of $3.16_{-0.35}^{+0.51}$~d. The upper bound on the rotational period from \vsini is $4.07^{+0.45}_{-0.38}$~d.

\section{Analysis and results}\label{section:analysis_results}

We performed a blind search for transit signals in the TESS data with \nuance \citep{Garcia2024}, and did not recover any additional significant transit signal beyond those of planets~b and~c. We therefore do not confirm the shallower transit of the third planet candidate proposed by \citet{Salinas2025}. A preliminary inspection of the TESS light curves revealed the presence of significant TTVs. We therefore fit the observed photometry using a photodynamical model \citep{Carter2011} that accounts for the gravitational interactions between the star and the two transiting planets. The positions of the star and planets were computed as a function of time using the n-body code \rebound \citep{Rein2012} with the \whf integrator \citep{Rein2015}. We adopted a time step of 0.02~d, which resulted in a maximum error \citep{Almenara2018} below 1~ppm for the photometric model. The sky-projected positions were used to generate the light curve with \batman \citep{Kreidberg2015}, including the light-time effect \citep{Irwin1952}. Oversampling factors of 15, 5, 3, and 3 were applied to the model for the TESS 1800~s, 600~s, 200~s, and 120~s cadence data, respectively, to account for the integration-induced distortion described by \citet{Kipping2010}. The oversampled model was then binned back to match the cadence of the observed data points. The model was parametrised using the stellar mass and radius, limb-darkening coefficients, planet-to-star mass and radius ratios, and Jacobi orbital elements (Table~\ref{table:results}) at the reference epoch ($t_{\mathrm{ref}}=2\,460\,122.4554$~BJD$_{\mathrm{TDB}}$). Due to the symmetry of the problem, we fixed the longitude of the ascending node of planet~c at $t_{\mathrm{ref}}$ to 180\degree\ and constrained the inclination of planet~b to be less than 90\degree. We modelled the error terms of the transit light curves using Gaussian process regression, adopting the approximate Matern kernel implemented in \celerite \citep{Foreman-Mackey2017}. We used different kernel hyperparameters for each photometry dataset, corresponding to each TESS sector, ASTEP light curve, and CHEOPS observation. For CHEOPS observations, we used the satellite roll angle as a regressor, and we added a linear model with a second-order term with log-background. In total, the model comprises 53 free parameters. We adopted normal priors for the stellar mass and radius from Sect.~\ref{section:stellar_parameters}, and uninformative priors for the remaining parameters. The joint posterior distribution was sampled using the \emcee algorithm \citep{Goodman2010,emcee}.

\begin{figure}
  \centering
  \includegraphics[width=0.48\textwidth]{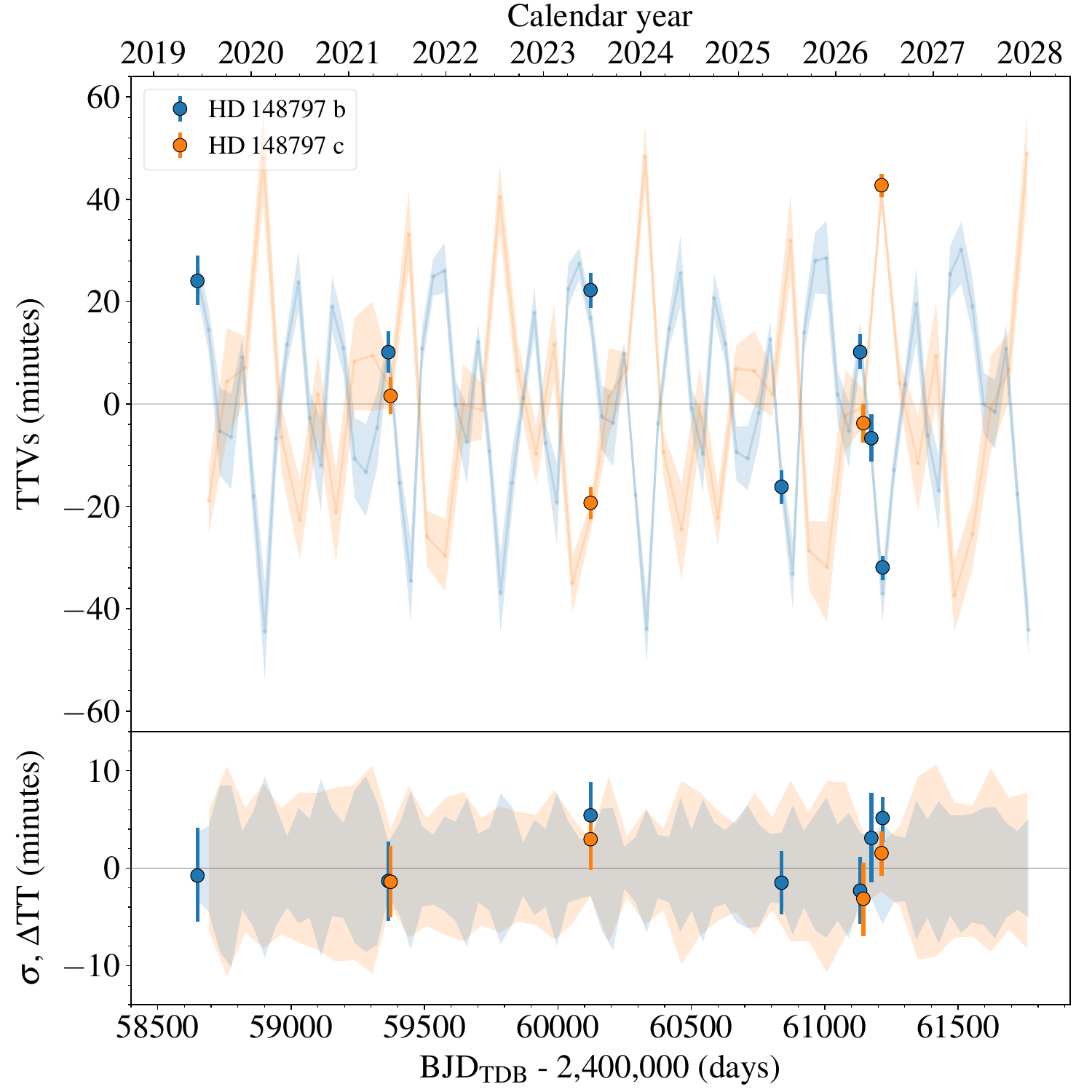}
  \caption{Posterior TTV predictions of planets\,b (blue band) and c (orange band) computed relative to a linear ephemeris (Table~\ref{table:results}). A thousand random draws from the posterior distribution were used to estimate the median TTV values and their uncertainties (68.3\% confidence interval). The upper panel shows the posterior TTV values and compares them with the individual transit-time determinations (Table~\ref{table:transit_times}, error bars). In the lower panel, the posterior median transit-timing value was subtracted to emphasize the uncertainty in the distribution and facilitate a clearer comparison with the individual transit times.} \label{figure:TTVs}
\end{figure}

Table~\ref{table:results} lists the median and the 68\% credible interval (CI) of the marginal distribution of the inferred system parameters. The MAP model for the transit photometry data is shown in Figs.~\ref{figure:phot}\footnote{Contrary to the possibility raised by \citet{Salinas2025}, if both planets orbit in the same direction, the double transit in TESS Sector 66 cannot be a mutual event \citep{Ragozzine2010}, because the inner planet started its transit first and crosses the stellar chord faster than the outer planet.} and \ref{figure:ASTEP}. Figure~\ref{figure:TTVs} presents the posterior TTVs for the two planets, derived from the times of minimum projected separation between the star and each planet, and compares them with individually determined transit times (Table~\ref{table:transit_times}). These individual transit times were computed with \juliet \citep{Espinoza2019,Kreidberg2015,Speagle2020}, assuming a constant transit duration and the same noise model as in the photodynamical analysis. The anticorrelated TTVs of HD\,148797~b and HD\,148797~c dynamically confirm that the two transit signals are produced by gravitationally interacting planets orbiting the same star. The photodynamical analysis yields masses of $39.3^{+13}_{-8.5}~\mathrm{M_E}$ and $39.6 \pm 9.3~\mathrm{M_E}$ for planets~b and~c, respectively, corresponding to mass constraints at the $4.6\sigma$ and $4.3\sigma$ levels.

Figure~\ref{figure:pyramid} shows the posterior distributions of the main inferred system parameters. The posterior is bimodal in the inclination of planet~c, reflecting the two possible transit geometries above and below $90^\circ$. The eccentricities of both planets are weakly constrained, with posterior values between $\sim$ 0 and 0.4. Because of the general mass--eccentricity correlation in TTV analyses, lower eccentricities would imply higher masses, while higher eccentricities would imply lower masses. Overall, additional transit observations of both planets would be the most direct way to further refine the orbital solution. An RV mass determination would be challenging for a $\vsini \sim 20$~\kms mid-F star, with expected semi-amplitudes of $6.2^{+2.0}_{-1.3}$ and $5.3 \pm 1.2$~\ms for planets~b and c, respectively. We note, however, that the Rossiter--McLaughlin effect \citep{Rossiter1924,McLaughlin1924}, which can be used to measure projected spin--orbit obliquities, should be more accessible, with expected amplitudes\footnote{Estimated as $\left(R_p/R_\star\right)^2 v\sin i_\star \sqrt{1-b^2}$.} of $\sim$22 and $\sim$32~\ms for planets~b and c, respectively.

\section{Discussion}\label{section:discussion}

\begin{figure*}[htb!]
    \centering
    \includegraphics[width=\textwidth]{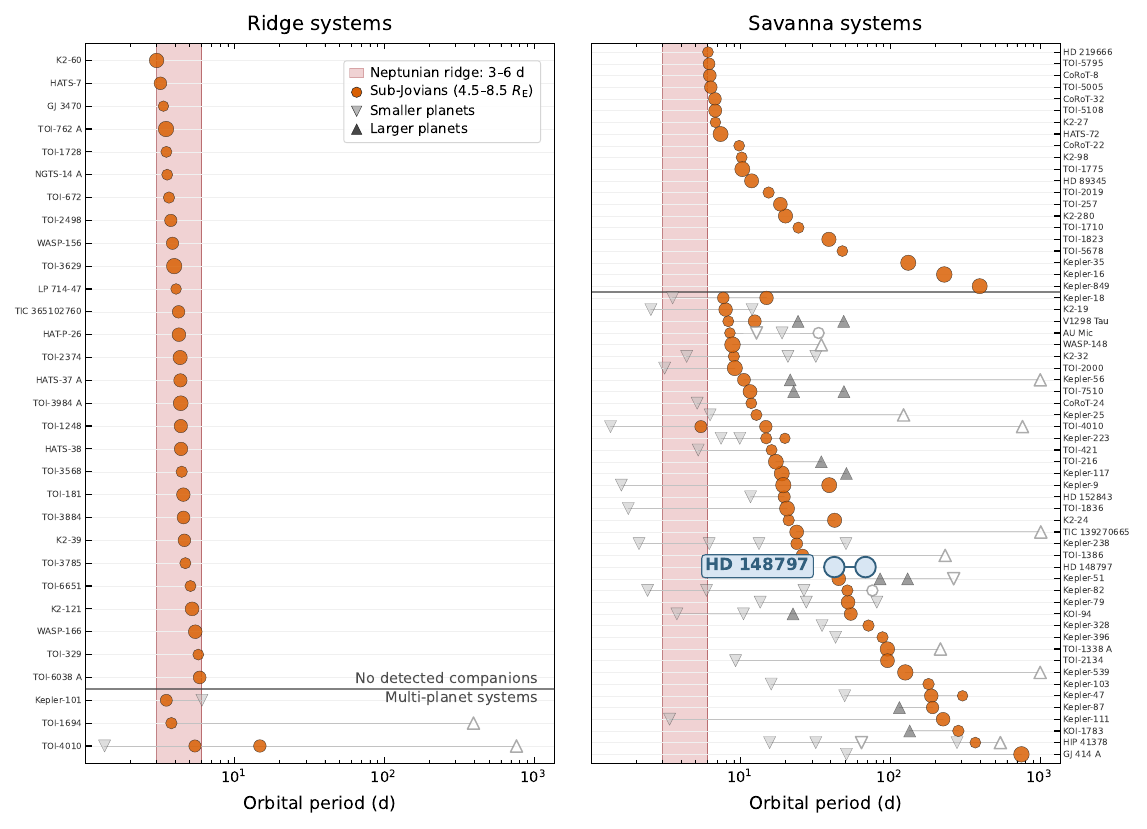}
    \caption{Detected orbital architectures of planetary systems in the Neptunian ridge (left panel) and savanna (right panel). The sample is drawn from the NASA Exoplanet Archive \citep{Christiansen2025} and includes systems hosting at least one planet with measured mass and radius in the range $4.5$--$8.5$~\REarth. Systems are grouped according to whether the selected sub-Jovian planet lies in the ridge ($3$--$6$~d; shaded region) or in the savanna ($P_{\rm orb}>6$~d) \citep{CastroGonzalez2024a}. Within each panel, systems are separated into those with and without detected companions, and ordered by the period of the selected planet in that region. All known companions in those systems are shown: orange circles denote planets in the selected radius range, while downward and upward triangles mark smaller and larger companions, respectively. Open symbols indicate planets with estimated radii.}
\label{fig:architecture}
\end{figure*}

HD\,148797 provides a valuable benchmark for the Neptunian savanna at moderate orbital periods, where the architectures, bulk properties, and dynamical evolution of sub-Jovian planets remain comparatively less explored than at shorter periods. We first place HD\,148797 in the context of detected savanna and ridge architectures (Sect.~\ref{section:architecture}) and of the period--density distribution of sub-Jovian planets (Sect.~\ref{section:bulk_densities}). We then discuss the TTV dynamics (Sect.~\ref{section:ttv_dynamics}), investigate the internal structures of the two planets (Sect.~\ref{section:internal_structure}), and assess their prospects for comparative atmospheric characterisation (Sect.~\ref{section:comparative_planetology}).

\subsection{Compact multi-planet architectures in the savanna}
\label{section:architecture}

HD\,148797 is composed of two transiting sub-Jovian planets at orbital periods of 42.1 and 68.2~d, corresponding to a period ratio of 1.619. The two planets have similar radii and masses, modest eccentricities, and a low mutual inclination, forming a compact configuration at moderate orbital periods. Such an architecture is difficult to reconcile with a strongly disruptive migration history and is more naturally compatible with dynamically cold evolution, including possible disk-driven migration from larger orbital distances. HD\,148797 therefore provides a useful test case for assessing whether similar compact configurations are common among savanna sub-Jovians.

Figure~\ref{fig:architecture} compares the detected orbital architectures of systems across the ridge and savanna, including HD\,148797. The fiducial sample is defined by selecting planets with radii between $4.5$ and $8.5$~\REarth and measured masses. For each selected system, we then show all reported planetary companions, irrespective of their radii or masses. We find that multi-planet architectures are substantially more commonly detected in the savanna than in the ridge: $66\%$ of savanna systems have at least one detected planetary companion ($40/61$), compared with $9.7\%$ of ridge systems ($3/31$). This contrast is also recovered when we broaden the selection to include RV-detected planets without measured radii, yielding detected multi-system fractions of $70\%$ in the savanna ($82/118$) and $22\%$ in the ridge ($8/37$). Notably, the savanna multi-planet systems are often compact: $75\%$ of them ($30/40$) contain at least one adjacent planet pair with $P_{\rm out}/P_{\rm in}<3$, and the median adjacent period ratio among planet pairs is 2.06.

\begin{figure}
    \centering
    \includegraphics[width=0.48\textwidth]{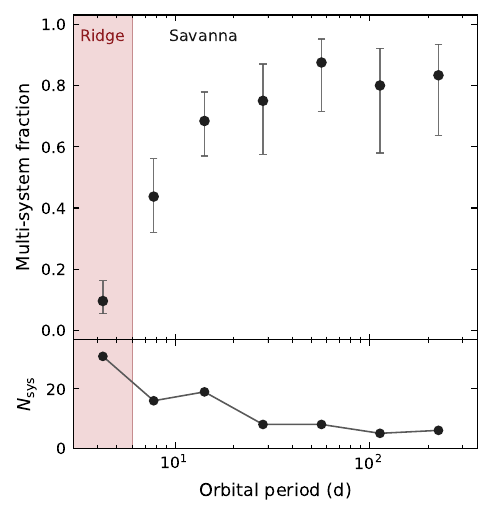}
    \caption{Observed multi-planet system fraction as a function of orbital period for sub-Jovian planets ($4.5$--$8.5$~\REarth) drawn from the NASA Exoplanet Archive \citep{Christiansen2025}. The shaded region marks the ridge ($\simeq 3$--$6$~d) and the remaining bins correspond to the savanna \citep{CastroGonzalez2024a}. The lower panel shows the number of systems in each period bin.}
\label{fig:multi_fraction}
\end{figure}

To examine whether the observed high incidence of multi-planet architectures persists across different regions of the savanna, Fig.~\ref{fig:multi_fraction} shows the multi-system fraction as a function of orbital period. The fraction remains high beyond $\sim10$~d and out to $\sim300$~d: typically $\sim70$--$90\%$ of systems in this region have at least one detected planetary companion. The innermost savanna bin ($6$--$10$~d) suggests a lower multi-system fraction, a drop that becomes clearly established in the ridge. Thus, the observed transition from isolated to compact multi-planet architectures occurs close to the ridge--savanna boundary, in an otherwise well-sampled region of parameter space. We caution, however, that while the high multiplicity of savanna systems is a direct observational result, the isolation of ridge systems is inferred from the absence of detected companions and should ultimately be tested with a homogeneous occurrence-rate analysis (e.g. Thomas et al., submitted). Nevertheless, the large contrast seen in the detected architectures, together with the analogy to the isolation of hot Jupiters in the pile-up, suggests a common qualitative picture in which the shortest-period Jovian and sub-Jovian overdensities are preferentially populated by isolated planets that underwent disruptive HEM. Future astrometric surveys (e.g. \textit{Gaia} DR4) may test this scenario by revealing the massive, long-period perturbers expected in a fraction of ridge and hot-Jupiter pile-up systems.

\subsection{Low-density prevalence in the moderate-period savanna}\label{section:bulk_densities}

\begin{figure*}
  \centering
  \includegraphics[width=0.98\textwidth]{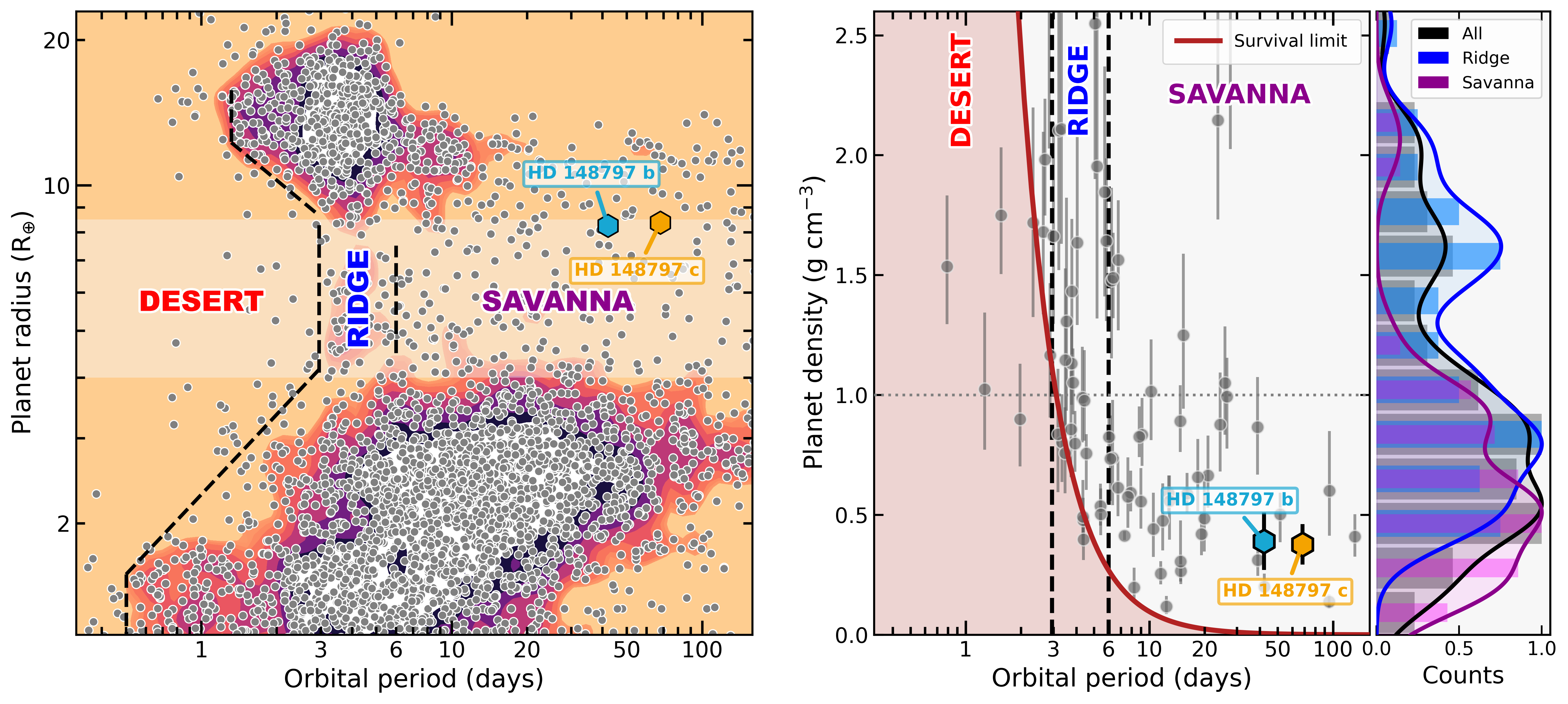}
  \caption{Location of HD\,148797~b and c in the exo-Neptunian landscape. Left: period--radius diagram of the known exoplanet population, showing the Neptunian desert, ridge, and savanna regions as defined by \citet{CastroGonzalez2024a}. Right: period--density diagram for the subset of planets with measured bulk densities. In the density plane, the red curve shows the tidal survival limit expected after HEM from \citet{CastroGonzalez2026}. This plot was produced with \href{https://github.com/castro-gzlz/nep-des}{\texttt{nep-des}}.}
  \label{figure:nep-des}
\end{figure*}

In Fig.~\ref{figure:nep-des}, we place HD\,148797 in the context of the known exoplanet population in the period--radius and period--density spaces. We focus on the regions corresponding to the Neptunian savanna, ridge, and desert, as defined by \citet{CastroGonzalez2024a}. The moderate orbital periods of the two HD\,148797 planets place them in a relatively sparsely populated region of the parameter space, particularly in the period--density plane, reflecting the observational difficulty of measuring precise masses for longer-period transiting planets. We use the newly measured masses and densities of HD\,148797~b and c to revisit the density trend reported by \citet{CastroGonzalez2024b}, which was originally defined for savanna planets with orbital periods up to $\simeq$30~d, and extend the period range out to $\simeq$100~d. The sub-Jovian population in this extended-period savanna remains dominated by low bulk densities, mostly below 1~\gcm, in contrast with the density-enhanced population in the ridge \citep{CastroGonzalez2024b,CastroGonzalez2026}. Within the current data set, we find no clear evidence for an additional dependence of bulk density on orbital period across the studied savanna regime. In this context, the bulk densities of both HD\,148797 planets, $\sim$0.4~\gcm, fall near the peak of the observed density distribution of savanna sub-Jovians.

This result highlights the role of systems such as HD\,148797 in mapping the density structure of the savanna beyond the period range previously explored. If the high-density sub-Jovians observed in the ridge were already intrinsically dense before migrating inward, analogous objects should eventually be found at longer orbital periods, closer to their likely formation region. The lack of such a population out to $\sim$100~d therefore shifts the search for potential dense ridge progenitors to wider orbits than those probed here. Extending this density census to still longer periods will be key to determining whether a ``missing'' population of dense sub-Jovians exists at larger separations, or whether the enhanced densities of ridge planets are instead mainly produced by post-formation evolutionary processes.

\subsection{TTVs for a system close to the golden mean}\label{section:ttv_dynamics}

Given the linear ephemeris values of the periods, $\bar P_b$ and $\bar P_c$ (Table~\ref{table:results}),
at $\bar P_c/\bar P_b=1.6188$ the period ratio of HD\,148797
is very close to the golden mean $\varphi=\ff{1}{2}(1+\sqrt{5})\simeq 1.6180$, a number which has
interesting properties in the context of three-body stability (Mardling \& Almenara, {\it in prep}).
While this is most likely a coincidence, it places the system close to the second and third-order commensurabilities 
5:3 and 8:5
and in between the first-order commensurabilities 2:1 and 3:2, resulting in each of these harmonics,
as well as the synodic harmonics 1:1 and 2:2,
contributing similarly to the TTVs of both planets. 

That information is available in more than one harmonic (in principle) gives one the ability to fully characterize 
the system given a long enough observing baseline and enough transits, in constrast to systems for which 
only one harmonic is available. The latter tends to be the case for systems close to (but not in) resonance and
usually results in mass-eccentricity degeneracy \citep{Boue2012,Lithwick2012},
while systems which are {\it inside} a resonance (or simply in the librating state) exhibit power in 
a second resonant harmonic, in addition to the superperiod $P_{\rm super}=(n/\bar P_b-n'/\bar P_c)^{-1}$
(where $n'=n+1$ for a first-order resonance), allowing for the masses to be measured independent
of the eccentricities \citep{Nesvorny2016,Almenara2025,Almenara2026}.

\begin{figure}
\centering
\includegraphics[width=0.5\textwidth]{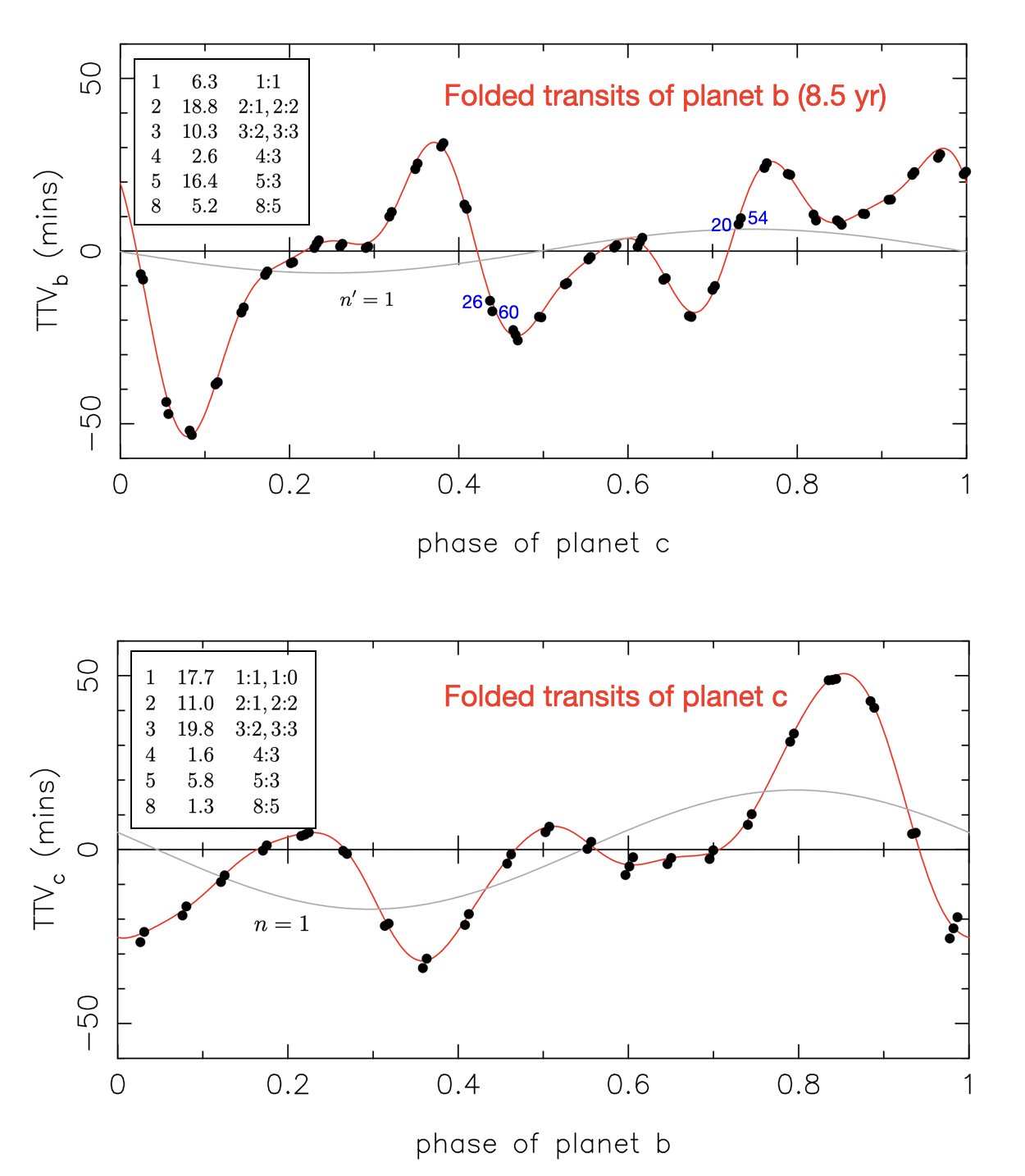}
\caption{Black dots: N-body MAP TTVs of planet b (bottom panel c),
folded at the linear ephemeris period of planet c (bottom panel b). 
Red curves are best Fourier fits, while grey curves show the $n'=1$ (top) and $n=1$ (bottom) harmonics explicitly. Inserts list Fourier harmonics and corresponding amplitudes in mins, and
the ``true'' contributing harmonics (see text for explanation). Harmonics not listed have amplitudes less than
1 min.
Bunching in sets of 2 and 3 is a result of proximity of the period ratio to 8/5,
while the fact that the bunches tend to be oblique to the best-fit curves is due to the slow secular evolution
of the longitudes of periastron on a timescale of around 600 yr (Mardling \& Almenara, {\it in prep}).
Blue numbers in the top panel are transit numbers (see main text).
}
\label{figure:fold}
\end{figure}

While the TTV time series of systems near or in resonance show clear sinusoidal behaviour, 
this is no longer the case for a system like HD\,148797 as is clear in Fig.~\ref{figure:TTVs}.
However, as demonstrated by \citealt{Almenara2015}
for the system Kepler-117 (for which the period ratio is 2.7), the stroboscopic
nature of transit timing information allows one to fold the inner planet's TTVs at the outer orbital period
(and the outer planet's TTVs at the inner orbital period)
to reveal a smooth curve which can be Fourier decomposed into several harmonics of the outer (inner)
orbital frequency, in spite of the fact that the true contributing harmonic frequencies are not commensurate.
Figure~\ref{figure:fold}

shows this behaviour for the TTVs of HD\,148797b and c, both of which contain significant power in several harmonics.
Here we give a brief overview of why one obtains such curves for systems away from resonance,
with a full analysis in Mardling \& Almenara ({\it in prep}).
Note that \citet{Deck2015} were the first to derive expressions \rn{ttvb} and \rn{ttvc} below,
but their focus was on systems close to (but outside of) 
resonance and in particular on the synodic part of the signal. Such systems
tend not to reveal a smooth curve on observing timescales when folded, precisely because they are close to a low-order
commensurability. Moreover, systems which are in the librating state will never visit all phases of 
the companion planet.

\begin{figure}
  \centering
  \includegraphics[width=0.47\textwidth]{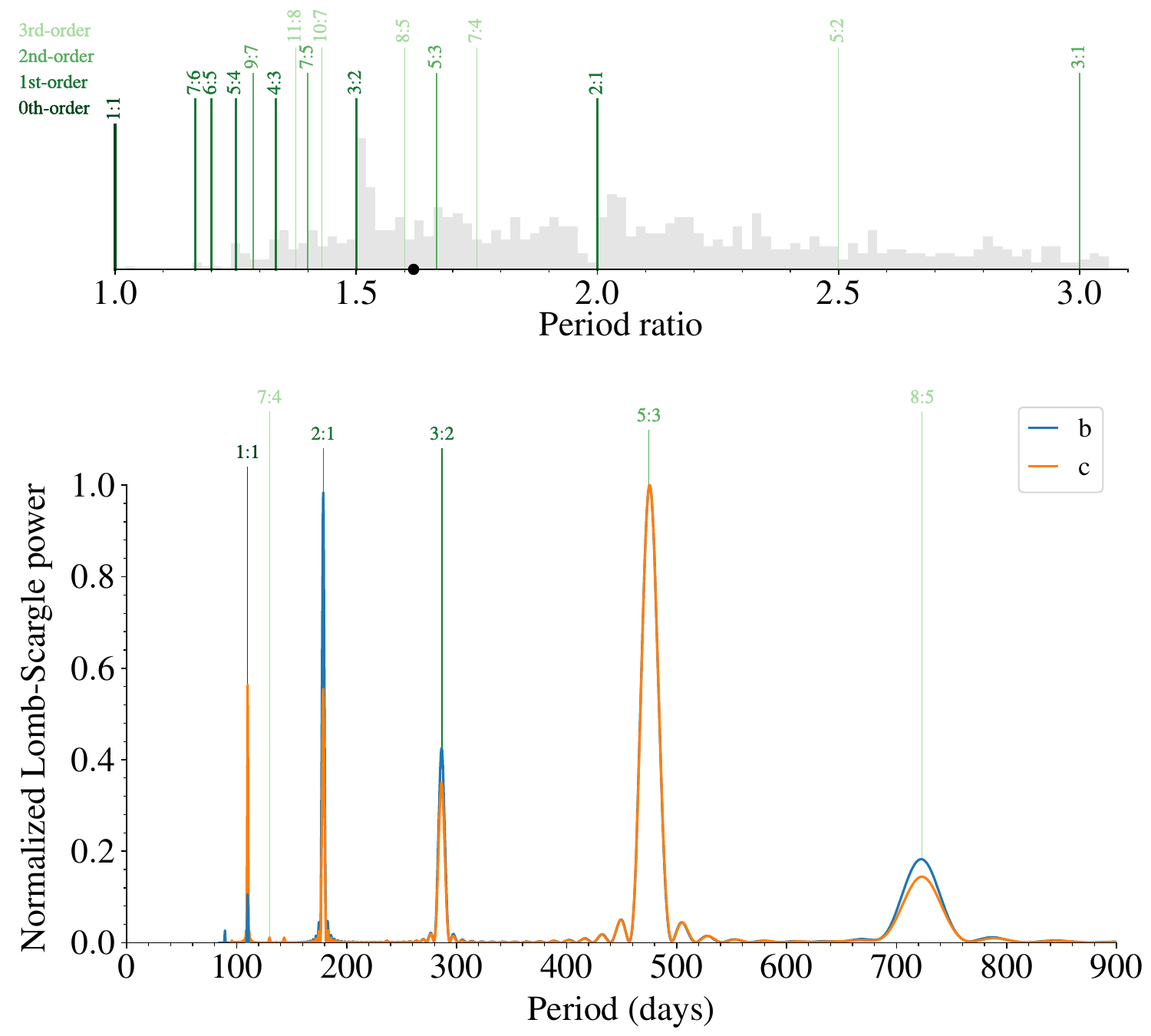}
  \caption{Lomb-Scargle periodogram \citep{Press1989} of the TTVs of planet~b (blue) and c (orange), computed from the MAP of the photodynamical model from the first transit observation up to the year 2050, showing the peaks associated with the modulation periods of different resonances. The upper panel displays the histogram of the period ratios of adjacent planets in known systems \citep[NASA Exoplanet Archive,][]{Christiansen2025}, including only planets whose orbital periods are measured to better than 2\%. The period ratio of planets~b and c (1.619) is indicated by a black dot on the $x$-axis. Vertical lines mark MMR of order zero to three. These are shown in the main panel when the corresponding superperiod lies between twice the period of planet~b and the maximum value of the $x$-axis.}\label{figure:LS}
\end{figure}

First note that TTVs occur at a fixed true longitude (for our coordinate system this is $3\pi/2$), 
while variations of the time between transits occur because the
orbits rotate (apsidal motion; for example, the time between transits is shorter when they occur near periastron), and exchange energy and angular momentum which manifests itself in the variation of the eccentricities and orbital
periods. Moreover, for a system away from resonance, the running means of the eccentricities, orbital periods 
and apsidal longitudes are approximately
constant over a typical observing baseline, while the TTVs capture shorter-period fluctuations
in these quantities due to variation of the mean longitude of the companion at the time of transit.
The rates of change of the elements can be obtained from a Fourier expansion of the 
interaction energy (disturbing function), for which the harmonic angles
involve the mean longitude combination 
\be
n\lambda_b-n'\lambda_c\simeq 2\pi(n/\bar P_b-n'/\bar P_c)(t-t_0)+n\lambda_b(t_0)-n'\lambda_c(t_0),
\ee
where $t_0$ is the time of the first transit. Variations of the elements obtained via integration
therefore involve the divisors $n/\bar P_b-n'/\bar P_c$ for the various $n':n$ harmonics, 
a quantity which is small when $\bar P_c/\bar P_b\simeq n'/n$.
The TTVs are then Fourier sums over these harmonics, but since (for transits of planet b) $\lambda_b(t_0)\simeq 3\pi/2$
and the $jth$ transit occurs at $t=t_j=j\bar P_b+t_0^{(b)}$, the TTVs of planet b take the form
\be
TTV_b(j)
\simeq
\frac{\bar P_b}{2\pi}\frac{m_c}{m_*}
\sum_{n'}
A_{n'}^{(b)}\sin(2\pi n'j/\sigma+\beta_{n'}^{(b)}),
\label{ttvb}
\ee
where $\sigma=\bar P_c/\bar P_b$, and the amplitude $A_{n'}^{(b)}$ and phase $\beta_{n'}^{(b)}$ are known
functions of $e_b$, $e_c$, $\alpha=a_b/a_c$ and the apsidal and outer mean longitude at $t=t_0^{(b)}$.
Similarly, the $kth$ transit of planet c occurs at $t=t_k=k\bar P_c+t_0^{(c)}$, so that the TTVs of planet c take the form
\be
TTV_c(k)
\simeq
\frac{\bar P_c}{2\pi}\frac{m_b}{m_*}
\sum_n
A_n^{(c)}\sin(2\pi nk\sigma+\beta_{n}^{(c)}).
\label{ttvc}
\ee
Therefore, even though the true contributing harmonics involve the (generally) non-commensurate 
frequencies $\nu=n\nu_b-n'\nu_c$, where $\nu_{b,c}\simeq 2\pi/\bar P_{b,c}$ are the mean motions,
the TTV signal itself can be decomposed into harmonics of the outer (inner) orbital frequency.
The true contributing frequencies are nonetheless revealed in a Lomb-Scargle of the TTV signal
(Fig.~\ref{figure:LS})
because this information is stored in the {\it spacing} of the TTVs on the folded curve, as is clear in 
Fig.~\ref{figure:fold} in which the proximity of the period ratio to the 8:5 commensurability causes the
TTVs to bunch together (in fact it takes twice the superperiod of the 8:5 harmonic to return, as indicated 
by the transit numbers shown in blue in the top panel of Fig.~\ref{figure:fold}
for which $34\bar P_b\simeq 1431$\,d, while
$P_{8:5}=(5/\bar P_b-8/\bar P_c)^{-1}\simeq 724$\,d).

With only seven transits of planet b and four of planet c, there are strong correlations between the 
planet masses and eccentricities, as is clear in the corner plot in Fig.~\ref{figure:pyramid}. However,
by carefully scheduling future transit observations, it is possible to optimize the power in the strongest harmonics,
and/or the harmonics which depend on the companion mass only (as is the case for the synodic $n:n$
harmonics),
thereby minimizing mass-eccentricity correlations.
The insets in both panels of Fig.~\ref{figure:fold} list the partial amplitudes of the harmonics of the folded curves,
along with the true contributing harmonics. For example, there is significant power in the synodic 1:1 true harmonic,
this having the shortest period at $P_{1:1}=(1/\bar P_b-1/\bar P_c)^{-1}\simeq 110$\,d. Moreover, for transits of b the
amplitude $A_1^{(b)}$ is independent of the eccentricities (at first order) and
the phase $\beta_1^{(b)}\simeq 0$ (see grey curve in top panel)
so that scheduling transits of planet b near times when $\lambda_b\simeq \pi/2$ and $3\pi/2$
will optimize the power in this harmonic of the signal.
On the other hand, for transits of the outer planet there is significant true power in the first-order 1:0 harmonic
(in addition to 1:1)
so that the $n=1$ harmonic has contributions from both. As a result the phase offset $\beta_1^{(c)}$ is slightly
non-zero (grey curve in bottom panel).
In the case of the $n'=2$ harmonic for transits of planet b, and the $n=2$ harmonic for transits of planet c, 
the 2:1 and 2:2 harmonics contribute similarly,
helping to separate the
masses from the eccentricities because the functional forms of the amplitudes are different.
These aspects will be made clear in our companion paper, where we will also discuss aliasing,
both from the point of view of contributing true harmonics such as 2:1 and 2:2, and also from
the point of view of the Nyquist frequency.

\subsection{Internal structure}\label{section:internal_structure}

\begin{figure}
  \centering
  \includegraphics[width=0.48\textwidth]{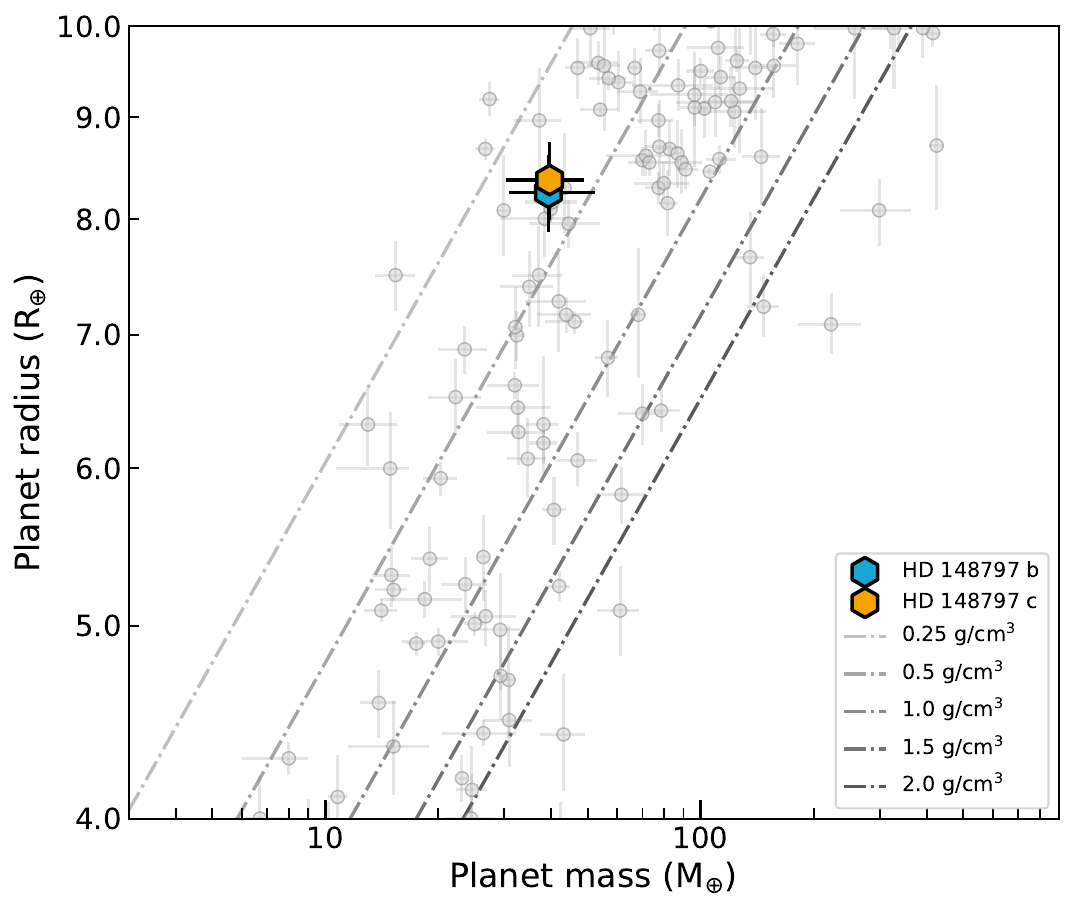}
  \caption{Mass--radius diagram for planets from the NASA Exoplanet Archive \citep{Christiansen2025} with masses and radii measured to better than 25\% and 10\%, respectively. Dashed lines represent iso-density curves. This plot was produced with \href{https://github.com/castro-gzlz/mr-plotter}{\texttt{mr-plotter}} \citep{CastroGonzalez2023}.}
  \label{figure:MR}
\end{figure}

Figure~\ref{figure:MR} places HD\,148797~b and c in the mass--radius diagram of well-characterized sub-Jovian planets, highlighting their nearly identical radii and comparable masses. This motivates a comparative assessment of their internal structures. We investigated these structures using \gastli\footnote{\url{https://github.com/lorenaacuna/GASTLI}} \citep{Acuna2021,Acuna2024}, a one-dimensional coupled interior-atmosphere model for warm ($T_{\rm eq}<1000$~K) giant planets with masses between 17~\MEarth and 6~\Mjup. The model assumes a two-layer structure with a core and an envelope. The core is represented by a 1:1 mixture of silicate rock and water, while the envelope is composed of H/He and water. The composition is described by two parameters: the core mass fraction (CMF) and the water mass fraction in the envelope ($Z_{\rm env}$), which is used as a proxy for the metal enrichment.

We generated model grids at equilibrium temperatures of 760~K for planet~b and 650~K for planet~c, assuming a solar C/O ratio of 0.55. The grids span planetary masses covering the photodynamical posterior in steps of 10~\MEarth, CMF values from 0 to 0.99 in steps of 0.1, $\log({\rm Fe/H})$ from $-2$ to 2.4 in steps of 0.5, and $T_{\rm int}$ from 50 to 250~K in steps of 25~K. For the retrieval, we used the planetary mass and radius, together with the isochronal stellar age, as observables, and sampled the posterior distribution with \emcee \citep{Goodman2010,emcee}, interpolating within the pre‑computed grids. The adopted priors and resulting posteriors are listed in Table~\ref{table:interior_retrieval}, and the one- and two-dimensional projections of the posteriors are shown in Fig.~\ref{figure:interior_retrieval}.

\begin{figure}
  \centering
  \includegraphics[width=0.47\textwidth]{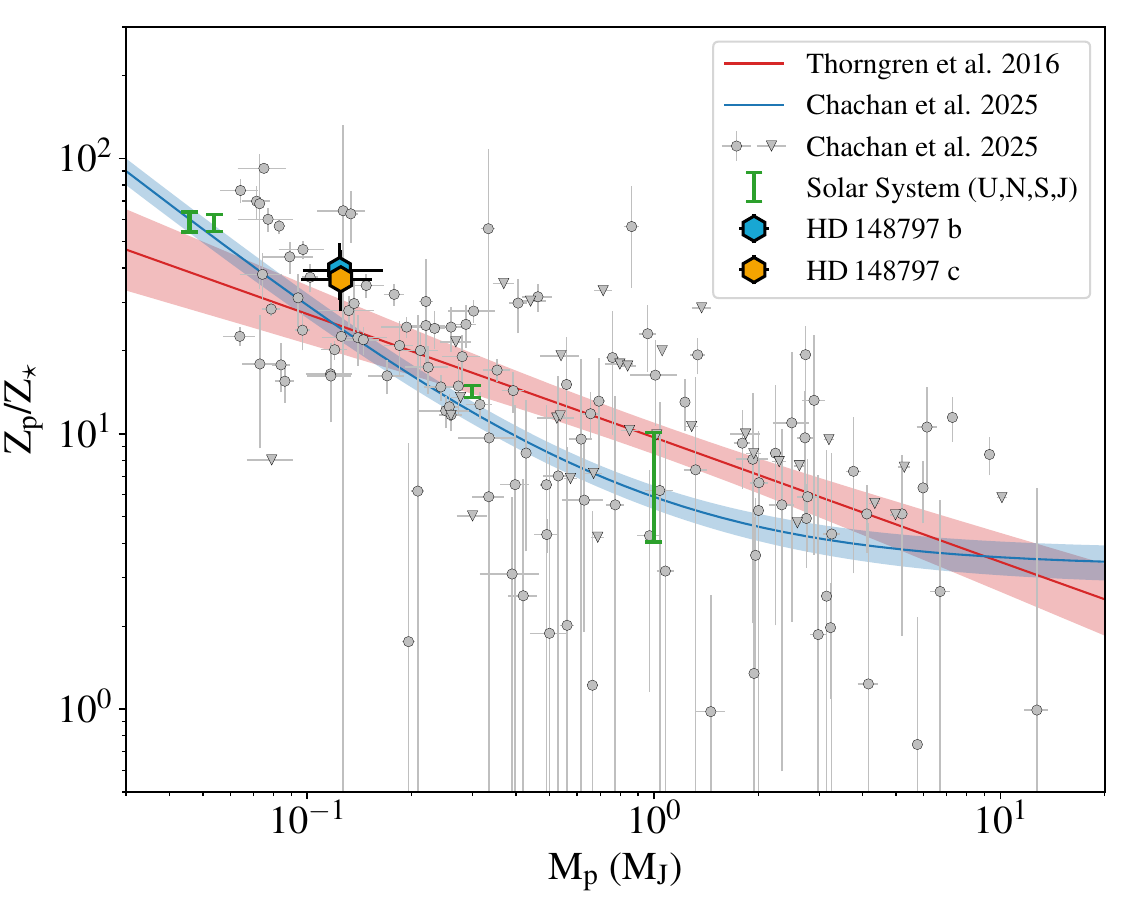}
  \caption{Heavy-element enrichment of non-inflated transiting planets relative to their parent stars as a function of mass. The grey error bars, with upper limits indicated by downward triangles, show the sample of planets analysed by \citet{Chachan2025}, and its fitted relation (median and 1$\sigma$ interval) is shown in blue. The relation from \citet{Thorngren2016} is shown in red. The range of values for Solar System giants \citep[][and references therein]{Miguel2023} is shown in green. The values for HD\,148797~b and c are plotted as a white-filled circles with black error bars.}\label{figure:Thorngren}
\end{figure}

We computed the stellar metal mass fraction as $Z_{\star}=0.014\times10^{\rm [Fe/H]_\star}$ and used it to infer the planetary heavy-element enrichment relative to the host star. This yields $Z_{\rm planet}/Z_{\star}=39.3^{+10}_{-8.6}$ for planet~b and $36.4^{+10}_{-8.4}$ for planet~c. In Fig.~\ref{figure:Thorngren}, we compare these values with the empirical mass--metal enrichment relations of \citet{Thorngren2016} and \citet{Chachan2025}. The inferred enrichments are broadly consistent with these trends, as expected for planets formed through core accretion.

The retrieved envelope mass fractions ($f_{\rm env} =1-\mathrm{CMF}$) of $0.55^{+0.11}_{-0.08}$ and $0.58^{+0.10}_{-0.08}$ place HD\,148797~b and c in an informative region of the sub-Jovian population. Planets with $f_{\rm env}\sim 50\%$ are expected to be rare as they should have started runaway gas accretion and grown to become gas giants in the classical core accretion model \citep{Mizuno1980,Stevenson1982,pollack1996}. \cite{Thomas2025} found that the distribution of $f_{\rm env}$ for a sample of 26 sub-Jovian planets was bimodal with a gap that might divide the sub-Jovian population into planets that did not reach the threshold for runaway gas accretion and those that started the process. However, the inferred $f_{\rm env}$ strongly depends on the atmospheric metallicity of the planets and as such the $f_{\rm env}$ gap moved depending on the assumed $\log({\rm Fe/H})$. Indeed, for HD\,148797~b and c we find the same degeneracy with significantly enriched envelopes ($\log({\rm Fe/H}) >>1$) leading to much smaller inferred core mass fractions (CMF$\leq0.2$). 
While the location of that gap in $f_\mathrm{env}$ depends on the assumed atmospheric metallicity, a corresponding gap in the hydrogen--helium mass fraction, at $f_\mathrm{H/He} \approx 0.3$, does not \citep[corroborating][]{Venturini2016}. With $f_\mathrm{H/He} = 1 - Z_\mathrm{planet} = 0.53 \pm 0.08$ and $0.56 \pm 0.08$, both HD\,148797 planets lie above this gap, in the population consistent with having crossed the runaway-accretion threshold. In this picture, HD\,148797~b and c would have entered runaway accretion only shortly before their gas supply was cut off, consistent with their moderate masses of $\sim$40\,$\mathrm{M_E}$. That two co-evolved planets in the same system halted at nearly indistinguishable envelope fractions is naturally explained if disk dispersal set a common endpoint to their accretion, in line with the dynamically cold history suggested by the system architecture (Sect.~\ref{section:architecture}).

\begin{table}[]
\scriptsize
\centering
\renewcommand{\arraystretch}{1.3}
\setlength{\tabcolsep}{5pt}
\caption{Interior structure retrieval of HD\,148797~b and c.}
\label{table:interior_retrieval}
\begin{tabular}{lccc}
\hline \hline
Parameter               & Prior & \multicolumn{2}{c}{Posterior median and 68.3\% CI}\\
& &            HD\,148797~b &   HD\,148797~c     \\ \hline
Core mass fraction, CMF                      & $U$(0.01, 0.99) & $0.454^{+0.077}_{-0.11}$   & $0.418^{+0.079}_{-0.10}$    \\
Atmospheric metallicity, $\log$(Fe/H)        & $U$(-2, 2.4)    & $-0.2 \pm 1.2$             & $-0.2 \pm 1.2$              \\
Envelope water mass fraction, $Z_{\rm env}$  &                 & $0.0121^{+0.097}_{-0.0081}$ & $0.0118^{+0.090}_{-0.0080}$ \\
Total water mass fraction, $Z_{\rm planet}$  &                 & $0.471 \pm 0.076$          & $0.437 \pm 0.076$           \\
Internal temperature, $T_{\rm int}$ ($K$)    & $U$(50, 250)    & $81^{+14}_{-10}$           & $78^{+13}_{-10}$            \\
\hline  
\end{tabular}
\tablefoot{$U$(l,u): Uniform distribution prior in the range [l, u].}
\end{table}

\subsection{Prospects for atmospheric characterisation}
\label{section:comparative_planetology}

\begin{figure}
  \centering
  \includegraphics[width=0.49\textwidth]{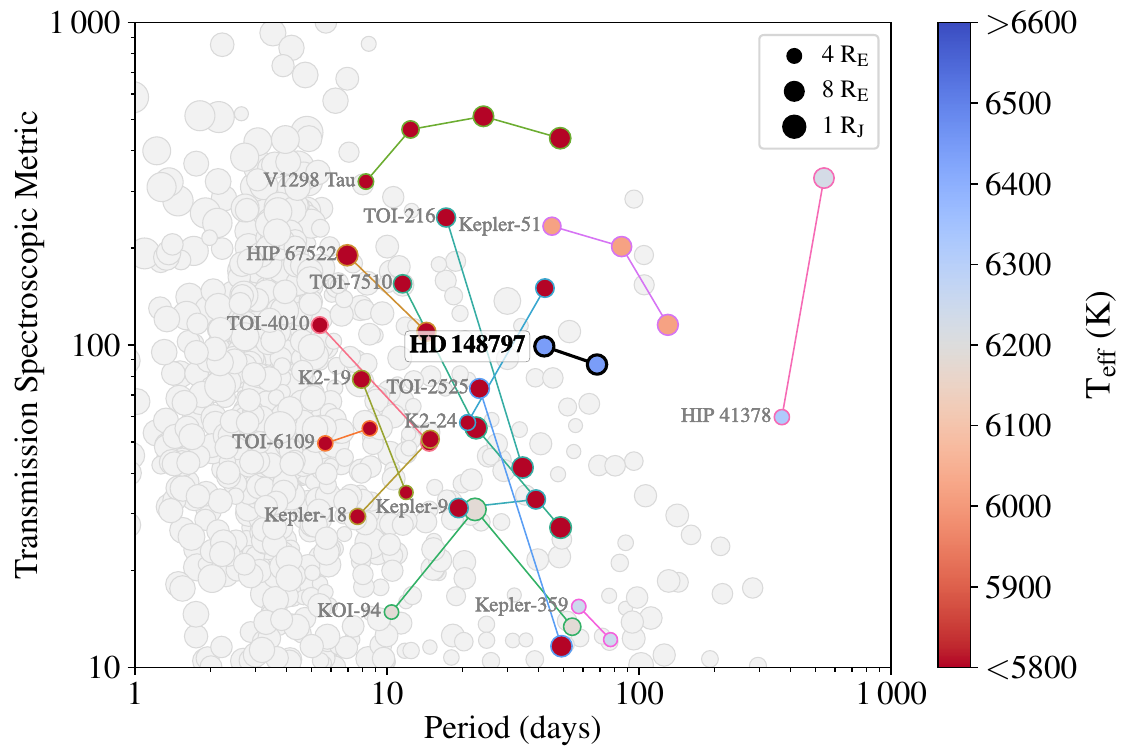}
  \caption{Transmission spectroscopy metric as a function of planetary orbital period for planets with $R_{\rm p}>4$~\REarth. Gray dots show the broader sample, with marker size proportional to planetary radius. Data are taken from the NASA Exoplanet Archive \citep{Christiansen2025}. Systems hosting two or more planets with $R_{\rm p}>4$~\REarth and TSM $>10$ are labelled. Each labelled system is shown in a distinct colour, and its planets are connected by a line matching the marker edge colour. The marker fill colour encodes the stellar effective temperature. The HD\,148797 planets are highlighted in black.}
\label{figure:TSM}
\end{figure}

With transmission spectroscopy metrics \citep[TSM;][]{Kempton2018} of $99 \pm 27$ for planet~b and $87 \pm 21$ for planet~c, both planets are promising sub-Jovian targets at moderate orbital periods for atmospheric characterisation. As shown in Fig.~\ref{figure:TSM}, this places HD\,148797 among the relatively few multi-planet systems hosting more than one planet with $R_{\rm p} > 4$~\REarth that are amenable to atmospheric follow-up. It also appears to be the only such system currently known in the warm Neptunian savanna.

The value of this configuration is that the two planets share the same host star and therefore the same stellar metallicity, age, and irradiation history. Atmospheric metallicity and C/O ratio are expected to encode aspects of a planet's formation location and accretion history \citep{Oberg2011,Madhusudhan2012}, but population-level comparisons are affected by differences between host stars and protoplanetary disks. In a co-evolved system such as HD\,148797, atmospheric differences between the two planets can instead be interpreted more directly in terms of their orbital locations, irradiation levels, and accretion histories. This makes the system a useful future target for comparative atmospheric studies of sub-Jovian planets.

The atmospheres of the HD\,148797 planets are within reach of current facilities. At these TSM values, JWST observations, for example with NIRISS/SOSS or NIRSpec/PRISM, could detect major molecular species such as H$_2$O, CO$_2$, and CH$_4$, and constrain the atmospheric metallicity and C/O ratio of each planet. The bright host may also make the system accessible to ground-based high-resolution cross-correlation spectroscopy \citep[e.g. CRIRES$^+$;][]{Brogi2019}. Although the moderate irradiation and orbital periods of HD\,148797~b and c make strong atmospheric escape less likely than for close-in gas giants, recent helium surveys of planets around F-type stars highlight the broader interest of this host-star regime for atmospheric evolution studies \citep[e.g.][]{2026AJ....171..257S}. Overall, atmospheric characterisation of this system would provide a direct test of whether the two planets share similar atmospheric compositions despite their different orbital periods and irradiation levels.

\section{Conclusions}
\label{section:conclusions}

We have confirmed and characterized two transiting sub-Jovian planets orbiting the bright F-type star HD\,148797. The planets have orbital periods of 42.1 and 68.2~d and radii of $8.25 \pm 0.37$ and $8.37 \pm 0.38$~\REarth, placing both objects in the moderate-period Neptunian savanna. Their significant anti-correlated TTVs dynamically confirm that the two signals arise from gravitationally interacting planets in the same system and yield masses of $39.3^{+13}_{-8.5}$ and $39.6 \pm 9.3$~\MEarth for planets~b and c, respectively. The resulting bulk densities, $0.39 \pm 0.12$ and $0.377 \pm 0.084$~\gcm, identify both planets as low-density sub-Jovians.

HD\,148797 provides a useful reference system for linking the bulk properties of savanna sub-Jovians to their orbital architectures. Its two planets form a compact, low-mutual-inclination architecture, a configuration that is more naturally compatible with dynamically cold evolution than with a strongly disruptive migration history. In the broader population, we find that compact multi-planet architectures are common in the savanna, with detected multi-system fractions of $\sim70$--$90\%$, and that most savanna multi-planet systems contain at least one adjacent planet pair with $P_{\rm out}/P_{\rm in}<3$.

The low densities of HD\,148797~b and c contribute to extending the observed low-density savanna population to longer orbital periods. We find no evidence, within the current sample, for a population of dense sub-Jovians out to $\sim100$~d analogous to the high-density planets observed in the ridge. This suggests that possible dense ridge progenitors, if present, may reside at wider orbits than those probed here, or, alternatively, that the enhanced densities of ridge planets are mainly produced by post-formation evolutionary processes. The bright host star, together with the similar radii and masses of the two planets, makes HD\,148797 a promising system for future comparative atmospheric studies of co-evolved savanna sub-Jovians.

\begin{acknowledgements}

This paper includes data collected by the TESS mission. Funding for the TESS mission is provided by the NASA's Science Mission Directorate.

Resources supporting this work were provided by the NASA High-End Computing (HEC) Program through the NASA Advanced Supercomputing (NAS) Division at Ames Research Center for the production of the SPOC data products.

This paper includes data collected by the TESS mission that are publicly available from the Mikulski Archive for Space Telescopes (MAST).

This work is based on observations collected with EulerCam mounted on the 1.2~m Swiss Euler telescope at La Silla Observatory, Chile.

This paper uses data obtained with the ASTEP telescope, at Concordia Station in Antarctica. ASTEP benefited from the support of the French and Italian polar agencies IPEV and PNRA in the framework of the Concordia station program, from OCA, INSU, ANR (EXTRASTEP) and ESA through the Science Faculty of the European Space Research and Technology Centre (ESTEC).
The Birmingham contribution to ASTEP is supported by the European Union's Horizon 2020 research and innovation programme (grant's agreement n$^{\circ}$ 803193/BEBOP), and from the Science and Technology Facilities Council (STFC; grant n$^\circ$ ST/S00193X/1, ST/W002582/1, and ST/Y001710/1). 

CHEOPS is an ESA mission in partnership with Switzerland with important contributions to the payload and the ground segment from Austria, Belgium, France, Germany, Hungary, Italy, Portugal, Spain, Sweden, and the United Kingdom. The CHEOPS Consortium would like to gratefully acknowledge the support received by all the agencies, offices, universities, and industries involved. Their flexibility and willingness to explore new approaches were essential to the success of this mission. CHEOPS data analysed in this article will be made available in the CHEOPS mission archive (\url{https://cheops.unige.ch/archive_browser/}).

Simulations in this paper made use of the \rebound code which can be downloaded freely at \url{http://github.com/hannorein/rebound}. 

This work made use of \texttt{TESS-cont} (\url{https://github.com/castro-gzlz/TESS-cont}), which also made use of \texttt{tpfplotter} \citep{Aller2020} and \texttt{TESS-PRF} \citep{Bell2022}.

This work made use of \texttt{mr-plotter} (available at \url{https://github.com/castro-gzlz/mr-plotter}).

This research made use of \texttt{nep-des} (available in \url{https://github.com/castro-gzlz/nep-des})

These simulations have been run on the {\it Bonsai} cluster kindly provided by the Observatoire de Gen\`eve.

A.L. and J.M.A. acknowledges support of the Swiss National Science Foundation under grant number TMSGI2\_211697.
M.L. acknowledges support of the Swiss National Science Foundation under grant number PCEFP2\_194576.

This work has been carried out within the framework of the NCCR PlanetS supported by the Swiss National Science Foundation under grant 51NF40\_205606.

\end{acknowledgements}

\bibliographystyle{aa}
\bibliography{HD148797}

\begin{appendix}
\FloatBarrier

\section{Additional figures and tables}






\begin{figure}
    \centering
    \includegraphics[width=0.47\textwidth]{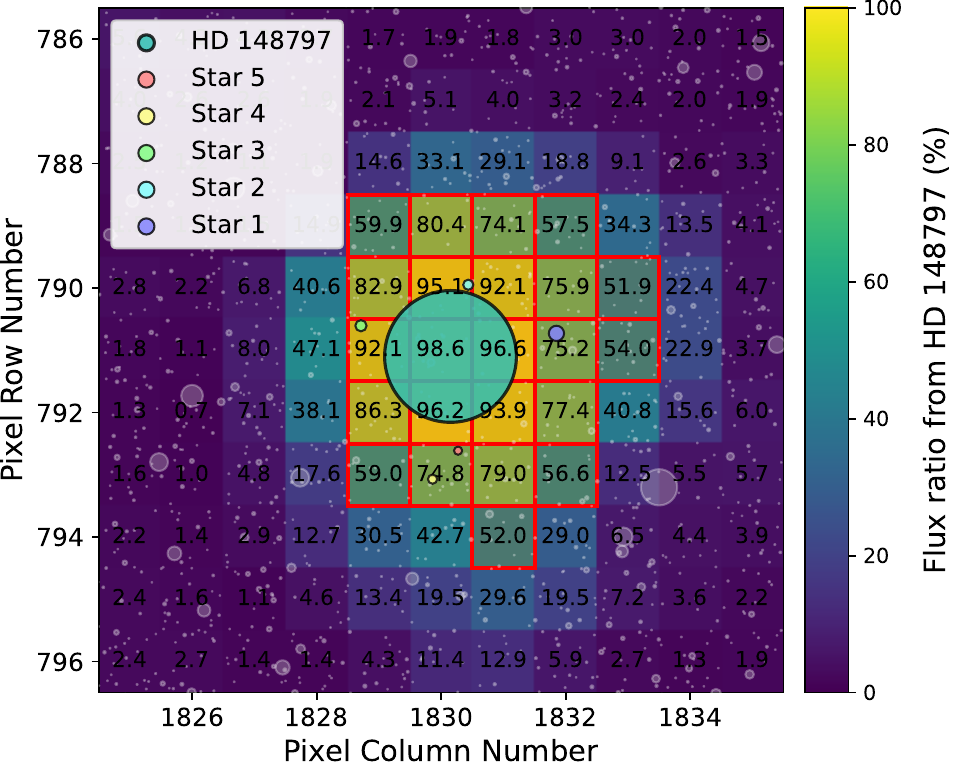}
    \caption{TPF-shaped heatmap of the pixel-by-pixel flux fraction from HD\,148797 in Sector~12. The red grid indicates the custom aperture used to extract the SAP photometry. White disks represent Gaia sources, with disk areas scaling with the emitted flux, while the five sources that contribute most to the aperture contamination are highlighted in different colours: Star~1, Star~2, Star~3, Star~4, and Star~5 contribute 0.52\%, 0.50\%, 0.35\%, 0.30\%, and 0.18\% of the aperture flux, respectively. This plot was created with \tesscont \citep{CastroGonzalez2024b}.}
    \label{fig:TESS-cont}
\end{figure}

\begin{table}[!ht]
    \scriptsize
    \renewcommand{\arraystretch}{1.25}
    \setlength{\tabcolsep}{2pt}
\centering
\caption{Modeling of the SED.}\label{table:SED}
\begin{tabular}{lccc}
\hline
Parameter & & Prior & Posterior median   \\
&  & & and 68.3\% CI \\
\hline
Effective temperature, $T_{\mathrm{eff}}$ & (K)     & $N$(6441, 51)     & $6441 \pm 50$ \\
Surface gravity, \logg\                   & (cgs)   & $N$(4.050, 0.077) & $4.147^{+0.017}_{-0.026}$ \\
Metallicity, $[\rm{Fe/H}]$                & (dex)   & $N$(-0.071, 0.079)& $-0.093 \pm 0.082$ \\
Distance                                  & (pc)    & $N$(171.77, 0.46) & $171.79 \pm 0.47$ \\
$E_{\mathrm{(B-V)}}$                      & (mag)   & $U$(0, 3)         & $0.031^{+0.028}_{-0.019}$ \\
Jitter \textit{Gaia}                      & (mag)   & $U$(0, 1)         & $0.083^{+0.11}_{-0.037}$ \\
Jitter 2MASS                              & (mag)   & $U$(0, 1)         & $0.031^{+0.057}_{-0.023}$ \\
Jitter WISE                               & (mag)   & $U$(0, 1)         & $0.061^{+0.073}_{-0.031}$ \\
Mass, $M_\star$                           & (M$_\odot$) &               & $1.268 \pm 0.048$ \\
Radius, $R_\star$                         & (R$_\odot$) &               & $1.569 \pm 0.022$ \\
Density, $\rho_\star$                     & ($\mathrm{g\;cm^{-3}}$) &   & $0.461 \pm 0.030$ \\
Isochronal age                            & (Gyr) &                     & $2.94 \pm 0.76$ \\
Luminosity                                & (L$_\odot$) &               & $3.82 \pm 0.14$ \smallskip\\
\hline
\end{tabular}
\tablefoot{$N$($\mu$,$\sigma$): Normal distribution prior with mean $\mu$, and standard deviation $\sigma$. $U$(l,u): Uniform distribution prior in the range [l, u].}
\end{table}

\begin{figure}[t]
  \centering
  \includegraphics[width=0.48\textwidth]{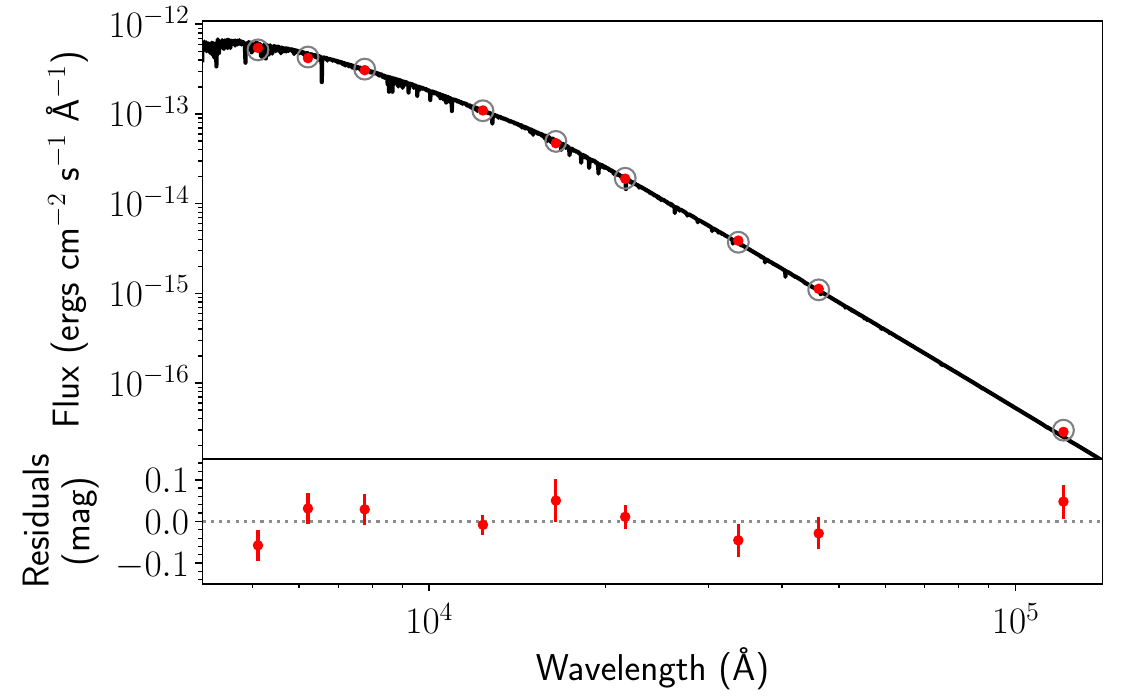}
  \caption{Spectral energy distribution of TOI-7510. The solid line shows the MAP PHOENIX/BT-Settl interpolated synthetic spectrum, red circles are the absolute photometric observations, and gray open circles are the result of integrating the synthetic spectrum in the observed bandpasses. The lower panel shows the residuals of the MAP model, with the jitter added quadratically to the data error bars.} \label{figure:SED}
\end{figure}

\begin{figure*}
  \centering
  \includegraphics[width=1\textwidth]{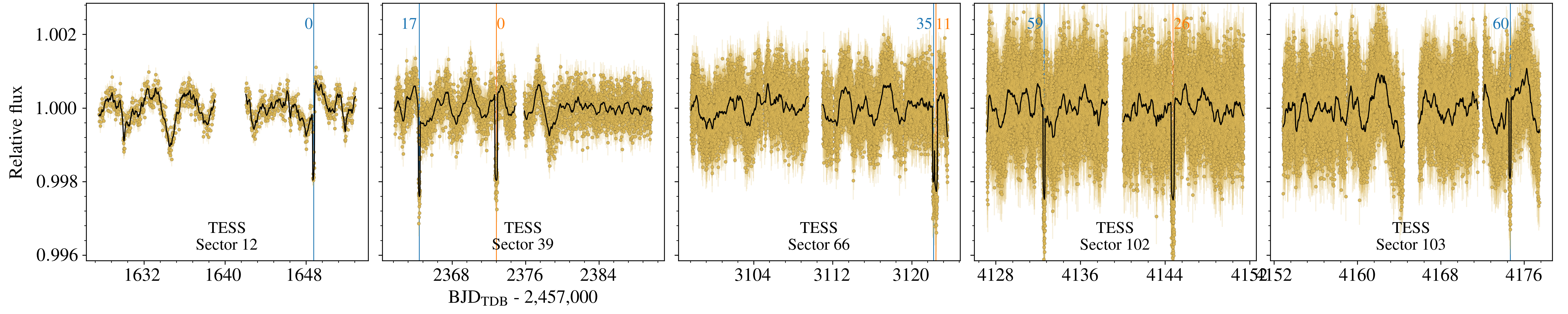}
  \caption{Modelling of the light curves of HD\,148797 observed by TESS in sectors 12, 39, 66, 102 and 103. The error bars represent the TESS observations and the black line shows the rotation kernel GP and transit model. Vertical lines mark the midtransit time of planets~b (blue) and c (orange) and are labelled by the number of orbital periods since the first observed transit.} \label{figure:Prot}
\end{figure*}

\begin{table*}
  \small
\renewcommand{\arraystretch}{1.1}
\setlength{\tabcolsep}{6pt}
\centering
\caption{Inferred system parameters.}\label{table:results}
\begin{tabular}{lccccc}
\hline
Parameter & Units & Prior & \multicolumn{2}{c}{Median and 68.3\% CI}  &  \\
\hline
\emph{\bf Star} \\
Stellar mass, $M_\star$              & (\Msun)     & $N(1.268, 0.080)$   & \multicolumn{2}{c}{$1.272 \pm 0.079$} \\
Stellar radius, $R_\star$            & (\Rnom)     & $N(1.569, 0.069)$   & \multicolumn{2}{c}{$1.547 \pm 0.060$} \\
Stellar mean density, $\rho_{\star}$ & ($\mathrm{g\;cm^{-3}}$) &         & \multicolumn{2}{c}{$0.485 \pm 0.068$} \\
Surface gravity, \logg\              & (cgs)       &                     & \multicolumn{2}{c}{$4.163 \pm 0.043$} \\
$q_1$ TESS, CHEOPS, ASTEP+ Blue, Red &             & $U(0, 1)$           & \multicolumn{2}{c}{$0.122^{+0.098}_{-0.063}$, $0.324 \pm 0.13$, $0.67^{+0.21}_{-0.31}$, $0.32^{+0.43}_{-0.28}$} \\
$q_2$ TESS, CHEOPS, ASTEP+ Blue, Red &             & $U(0, 1)$           & \multicolumn{2}{c}{$0.31^{+0.36}_{-0.23}$, $0.33 \pm 0.33$, $0.64^{+0.26}_{-0.37}$, $0.52 \pm 0.34$} \smallskip\\

\emph{\bf Planets} & &  & \emph{{\bf Planet~b}} & \emph{{\bf Planet~c}} \\
Semi-major axis, $a$                      & (au)                   &                                           & $0.2568 \pm 0.0052$         & $0.3536 \pm 0.0072$       \\
Eccentricity, $e$                         &                        &                                           & $0.153 \pm 0.093$           & $0.126 \pm 0.081$         \\
Argument of periastron, $\omega$          & (\degree)              &                                           & $207 \pm 20$                & $189 \pm 23$              \\
Inclination, $i$                          & (\degree)              & $S(0, 90)_{\rm b}, S(0, 180)_{\rm c}$     & $88.65 \pm 0.11$            & $89.115 \pm 0.070$ or $90.861 \pm 0.074$ \\
Longitude of the ascending node, $\Omega$ & (\degree)              & $U(90, 270)_{\rm b}$                      & $179.17 \pm 0.32$           & 180 (fixed at $t_{\mathrm{ref}}$)\\
Mean anomaly, $M_0$                       & (\degree)              &                                           & $263 \pm 27$                & $280 \pm 21$              \\
$\sqrt{e}\cos{\omega}$                    &                        & $U(-1, 1)$                                & $-0.32 \pm 0.13$            & $-0.33 \pm 0.11$          \\
$\sqrt{e}\sin{\omega}$                    &                        & $U(-1, 1)$                                & $-0.18 \pm 0.13$            & $-0.06 \pm 0.13$          \\
Mass ratio, $M_{\mathrm{p}}/M_\star$      &                        & $U(0, 1)$                                 & $\left(9.3^{+3.1}_{-2.0}\right)\e{-5}$ & $(9.4 \pm 2.2)\e{-5}$     \\
Radius ratio, $R_{\mathrm{p}}/R_\star$    &                        & $U(0, 1)$                                 & $0.04891 \pm 0.00091$       & $0.04956 \pm 0.00080$     \\
Scaled semi-major axis, $a/R_{\star}$     &                        &                                           & $35.7 \pm 1.6$              & $49.2 \pm 2.2$            \\
Impact parameter, $b$                     &                        &                                           & $0.877 \pm 0.015$           & $0.748 \pm 0.028$         \\
$T_0'$\;-\;2\;400\;000                    & (BJD$_{\mathrm{TDB}}$) & $U(59122, 61122)$                         & $60122.1984 \pm 0.0019$     & $60122.4509 \pm 0.0021$   \\
$P'$                                      & (d)                    & $U(0, 1000)$                              & $42.146 \pm 0.015$          & $68.100^{+0.019}_{-0.028}$ \\
$K'$                                      & (\ms)                  &                                           & $6.2^{+2.0}_{-1.3}$         & $5.3 \pm 1.2$             \\
Planet mass, $M_{\mathrm{p}}$             &(\MEarth)               &                                           & $39.3^{+13}_{-8.5}$         & $39.6 \pm 9.3$            \\
                                          &(\Mjup)                 &                                           & $0.124^{+0.041}_{-0.027}$   & $0.125 \pm 0.029$         \\
Planet radius, $R_{\mathrm{p}}$           &(\Renom)                &                                           & $8.25 \pm 0.37$             & $8.37 \pm 0.38$           \\
                                          &(\RJnom)                &                                           & $0.736 \pm 0.033$           & $0.746 \pm 0.034$         \\
Planet mean density, $\rho_{\mathrm{p}}$  &($\mathrm{g\;cm^{-3}}$) &                                           & $0.39 \pm 0.12$             & $0.377 \pm 0.084$         \\
Planet surface gravity, $\log$\,$g_{\mathrm{p}}$ &(cgs)            &                                           & $2.76 \pm 0.12$             & $2.748 \pm 0.092$         \\
Equilibrium temperature, T$_{\rm eq}$     & (K)                    &                                           & $762 \pm 17$                & $650 \pm 15$              \\

Mutual inclination, $I_{b,c}$             & (\degree)              &   &  \multicolumn{2}{c}{$0.91 \pm 0.25$ or $2.38 \pm 0.21$}  \smallskip\\

\emph{{\bf Linear ephemerides}} \\
Period                  & (d)                    & & 42.0988331  & 68.1514539   \\
$T_0$\;-\;2\;400\;000   & (BJD$_{\mathrm{TDB}}$) & & 58648.728195 & 58691.286130  \\

\hline
\end{tabular}
\tablefoot{\tiny The table lists: Prior, posterior median, and 68.3\% credible interval (CI) for the photodynamical analysis (Sect.~\ref{section:analysis_results}). The Jacobi orbital elements are given for the reference time $t_{\mathrm{ref}}=2\,460\,122.4554$~BJD$_{\mathrm{TDB}}$. The parameters $q_1$ and $q_2$ are the quadratic limb-darkening coefficients parametrised using \citet{kipping2013}. The planetary equilibrium temperature is computed for zero albedo and full day-night heat redistribution. $P'$ and $T_0'$ should not be confused with the linear ephemeris, and they were only used to reduce the correlations between jump parameters, replacing the semi-major axis and the mean anomaly at $t_{\mathrm{ref}}$. The linear ephemerides are derived from the median posterior transit times spanning the years 2019 to 2027. \\$T'_0 \equiv t_{\mathrm{ref}} - \frac{P'}{2\pi}\left(M_0-E+e\sin{E}\right)$ with $E=2\arctan{\left\{\sqrt{\frac{1-e}{1+e}}\tan{\left[\frac{1}{2}\left(\frac{\pi}{2}-\omega\right)\right]}\right\}}$, $P' \equiv \sqrt{\frac{4\pi^2a^{3}}{\mathcal G M_{\star}}}$, $K' \equiv \frac{M_p \sin{i}}{M_\star^{2/3}\sqrt{1-e^2}}\left(\frac{2 \pi \mathcal G}{P'}\right)^{1/3}$. CODATA 2018: $\mathcal G = 6.674\,30$\ten[-11]$\rm{m^3\,kg^{-1}\,s^{-2}}$. IAU 2012: au = $149\,597\,870\,700$~m$\,$. IAU 2015: \Rnom = 6.957\ten[8]~m, \GMnom = 1.327$\,$124$\,$4\ten[20]~$\rm{m^3\,s^{-2}}$, \Renom~=~6.378$\,$1\ten[6]~m, \GMenom = 3.986$\,$004\ten[14]~$\rm{m^3\,s^{-2}}$, \RJnom~=~7.149$\;$2\ten[7]~m, \GMJnom = 1.266$\;$865$\;$3\ten[17]~$\rm{m^3\;s^{-2}}$. \Msun$ = \GMnom/\mathcal G$, \MEarth = \GMenom/$\mathcal G$, \Mjup = \GMJnom/$\mathcal G$, $k^2$ = \GMnom$\,(86\,400~\rm{s})^2$/$\rm{au}^3$. $N(\mu, \sigma)$: Normal distribution with mean $\mu$ and standard deviation $\sigma$. $U(a, b)$: A uniform distribution defined between a lower $a$ and upper $b$ limit. $S(a, b)$: A sinusoidal distribution defined between a lower $a$ and upper $b$ limit.}
\end{table*}

\begin{table}[!b]
\centering
    \tiny
    \renewcommand{\arraystretch}{1.25}
    \setlength{\tabcolsep}{2pt}
\caption{Transit times of the observations.}\label{table:transit_times}
\begin{tabular}{rlll}
\hline
Epoch & Posterior median  & Telescope  \\
      & and 68.3\% CI [BJD$_{\mathrm{TDB}}$]  & Instrument  & \\
\hline
\emph{\bf Planet~b}\\ 
0  & $2458648.7449 \pm 0.0034$          & TESS Sector 12 \\
17 & $2459364.4154 \pm 0.0028$          & TESS Sector 39 \\
35 & $2460122.2028 \pm 0.0024$          & TESS Sector 66 \\
52 & $2460837.8563 \pm 0.0023$          & TESS Sector 93 \\
59 & $2461132.5664 \pm 0.0024$           & TESS Sector 102 \\
60 & $2461174.6535 \pm 0.0032$          & TESS Sector 103 \\
61 & $2461216.7349_{-0.0017}^{+0.0015}$ & CHEOPS, ASTEP+ Blue \& Red \smallskip\\

\emph{\bf Planet~c}\\ 
0  & $2459372.8018 \pm 0.0026$          & TESS Sector 39 \\    
11 & $2460122.4533 \pm 0.0022$          & TESS Sector 66 \\    
26 & $2461144.7359 \pm 0.0027$          & TESS Sector 102 \\      
27 & $2461212.9196 \pm 0.0016$          & CHEOPS, ASTEP+ Red  \smallskip\\
\hline
\end{tabular}
\end{table}

\begin{figure*}
  \hspace{-2cm}\includegraphics[width=1.2\textwidth]{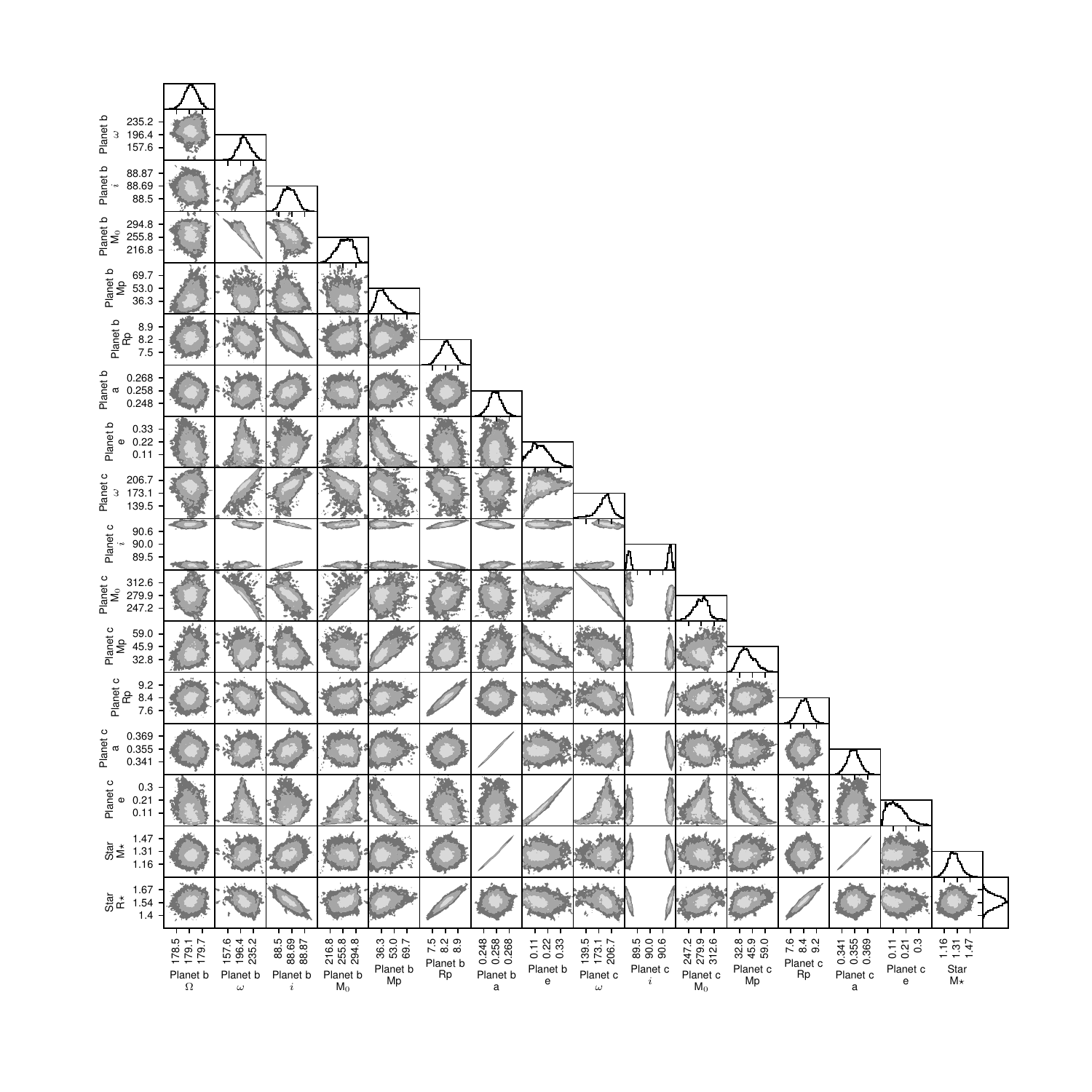}
  \vspace{-2cm}\caption{Two-parameter joint posterior distributions for the most relevant model parameters from the photodynamical modelling (Sect.~\ref{section:analysis_results}). The 39.3, 86.5, and 98.9\% two-variable joint confidence regions are denoted by three different gray levels; in the case of a Gaussian posterior, these regions project on to the one-dimensional 1, 2, and $3\,\sigma$ intervals. The histogram of the marginal distribution for each parameter is shown at the top of each column, except for the parameter on the last line, which is shown at the end of the line. Units are the same as in Table~\ref{table:results}.} \label{figure:pyramid}
\end{figure*}

\begin{figure*}[!ht]
  \includegraphics[width=0.49\textwidth]{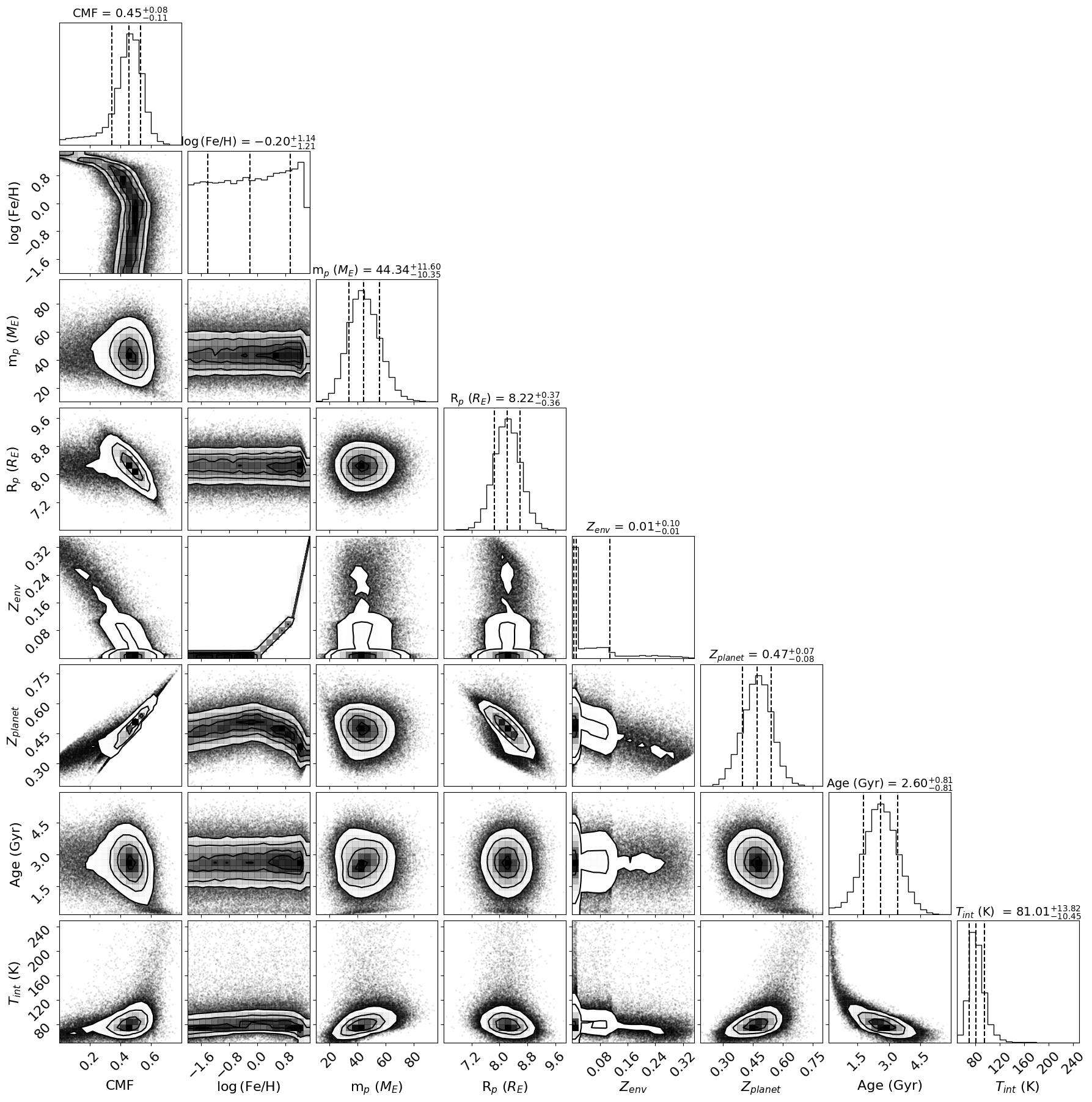}
  \includegraphics[width=0.49\textwidth]{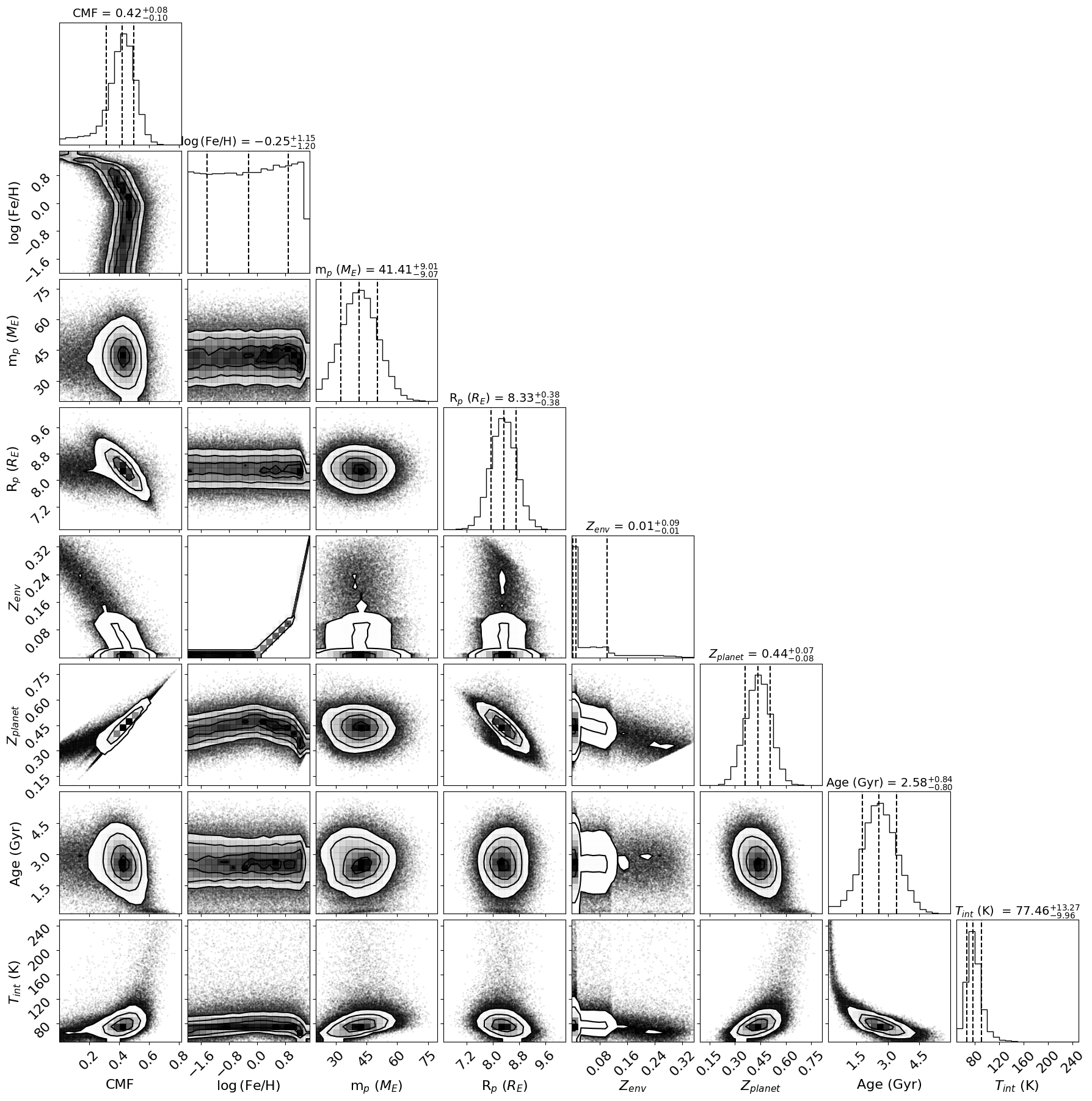}
  \caption{Two-parameter joint posterior distributions for the most relevant model parameters from the interior retrieval (Sect.~\ref{section:internal_structure}) of planet~b (left) and c (right). The 1, 2, and $3\,\sigma$ confidence regions are denoted black contours. The histogram of the marginal distribution for each parameter is shown at the top of each column, except for the parameter on the last line, which is shown at the end of the line.} \label{figure:interior_retrieval}
\end{figure*}

\end{appendix}
\end{document}